\documentclass[5p,times]{elsarticle} 

\usepackage{amssymb}

\usepackage{lineno}

\usepackage[T1]{fontenc} 
\usepackage{lmodern} 
\usepackage[final]{microtype} 
\usepackage{amsmath} 
\usepackage{booktabs} 
\usepackage{url} 
\usepackage{alltt} 
\usepackage{subfig} 
\usepackage[linesnumbered,ruled,vlined]{algorithm2e} 
\usepackage{rotating} 
\usepackage{xcolor} 

\def\imagecentered#1{$\vcenter{\null\hbox{#1}}$}

\newcommand{\martinezFigLabel}[6]{
  \begin{tabular}{@{}cc@{}}
    \begin{minipage}{0cm}
      \begin{sideways} \footnotesize #5~#6 \end{sideways}
    \end{minipage}
    &
    \imagecentered{\includegraphics[width=#1]{#2}}
    \tabularnewline

    &
    \footnotesize #3~#4
    \tabularnewline
  \end{tabular}
}

\newcommand{\teal}[1]{{\color{teal}{#1}}}
\newcommand{\purple}[1]{{\color{purple}{#1}}}
\newcommand{\brown}[1]{{\color{brown}{#1}}}
\newcommand{\violet}[1]{{\color{violet}{#1}}}
\newcommand{\gray}[1]{{\color{gray}{#1}}}

\newcommand{\pmf}[1]{#1}                  

\journal{Future Generation Computer Systems}

\begin{document}

\begin{frontmatter}



\title{Distributed and heterogeneous tensor--vector contraction algorithms for high performance computing\tnoteref{copyright,doi}}

\tnotetext[copyright]{\textcopyright 2025. This manuscript version is made available under the CC BY-NC-ND 4.0 license \url{https://creativecommons.org/licenses/by-nc-nd/4.0/}}
\tnotetext[doi]{Published journal article available at \url{https://doi.org/10.1016/j.future.2024.107698}}


\author[label1,label2]{Pedro J.~Martinez-Ferrer\corref{cor1}}
\ead{pedro.martinez.ferrer@upc.edu}

\affiliation[label1]{organization={Departament d'Arquitectura
            de Computadors~(DAC), Universitat Polit\`ecnica de Catalunya -
            BarcelonaTech~(UPC)},
            addressline={\\Campus~Nord, Edif.~D6, C.~Jordi~Girona~1-3},
            city={Barcelona},
            postcode={08034},
            state={Catalu\~{n}a},
            country={Spain}}

\affiliation[label2]{organization={Barcelona~Supercomputing~Center~(BSC)},
            addressline={Pl.~Eusebi G\"uell~\mbox{1-3}},
            city={Barcelona},
            postcode={08034},
            state={Catalu\~{n}a},
            country={Spain}}

\author[]{Albert-Jan Yzelman}
\ead{albert-jan@yzelman.net}


\author[label2]{Vicen\c{c} Beltran}
\ead{vbeltran@bsc.es}

\cortext[cor1]{Corresponding author}

\begin{abstract}
The tensor--vector contraction (TVC) is the most memory-bound
operation of its class and a core component of the higher\pmf{-}order
power method (HOPM).  This paper brings distributed-memory
parallelization to a native TVC algorithm for dense tensors that
overall remains oblivious to contraction mode, tensor splitting and
tensor order.  Similarly, we propose a novel distributed HOPM, namely
dHOPM$_3$, that can save up to one order of magnitude of streamed
memory and is about twice as costly in terms of data movement as a
distributed TVC operation (dTVC) when using task-based
parallelization.  The numerical experiments carried out in this work
on three different architectures featuring multi-core and
\pmf{accelerators} confirm that the performance\pmf{s} of dTVC and
dHOPM$_3$ remain relatively close to the peak system memory bandwidth
(50\%--80\%, depending on the architecture) and on par with STREAM
\pmf{benchmark figures}.  On strong scalability scenarios, our native
multi-core implementations of these two algorithms can achieve similar
and sometimes even greater performance figures than those based upon
state-of-the-art CUDA batched kernels.  Finally, we demonstrate that
both computation and communication can benefit from mixed precision
arithmetic also in cases where the hardware does not support low
precision data types natively.
\end{abstract}




\end{frontmatter}



\section{Introduction}\label{sec:introduction}
Tensors can be considered as multidimensional arrays that store data
in a certain manner according to multiple (i.e., multilinear)
attributes.  Exploiting them is of great importance since it allows to
extract patterns inherently present in such datasets.  It is the field
of multilinear algebra that defines tensor operations and algorithms
such as products, transformations, and decompositions to name but a
few~\cite{Kolda2009}.  A major kernel is the tensor contraction, which
includes the following operations: (i) tensor--tensor contraction
(TTC), (ii) tensor--matrix contraction (TMC), and (iii) tensor--vector
contraction (TVC), being all of them core components in widely used
tensor algorithms: the Khatri-Rao product~\cite{Khatri1968}, the
higher\pmf{-}order orthogonal iteration (HOOI)
algorithm~\cite{Lathauwer2000} that computes the truncated Tucker
decomposition~\cite{Tucker1966}, the alternating least squares (ALS)
algorithm for the canonical polyadic decomposition
(CPD)~\cite{Harshman1970}, or the higher\pmf{-}order power method
(HOPM)~\cite{Lathauwer2000}.

While TMC or TTC imply the contraction over two or more modes of a
given input tensor, TVC performs a contraction over one single mode,
rendering it a true bandwidth-bounded kernel.  In fact, TVC attains a
mere arithmetic intensity between 1 and 2~FLOP/byte.  Therefore, it
comes as no surprise that the TVC performance is often measured in
terms of memory bandwidth (GB/s) while, in the case of TMC and TTC
operations, the throughput is typically given in GFLOP/s, in analogy
to what occurs with the equivalent linear algebraic operations:
matrix--vector multiplication (MVM) and matrix--matrix multiplication
(MMM), respectively.  This contribution focuses exclusively on the
performance of the TVC on distributed systems and a more complex
tensor algorithm such as the HOPM.

The actual computation inside a TVC kernel can be achieved in
different ways, e.g.\ via the ``looped'' and ``unfolded'' algorithms.
Both of them appeal to the use of MVM operations, which requires the
conversion of the input tensor into its matricized form before the
actual contraction takes place.  While the looped variant recognizes
the input tensor as a series of contiguous matrices and carries out
several MVMs, the unfolded counterpart first needs to reorganize the
tensor in memory, at the expense of additional data movement, in order
to assemble a single large matrix and conduct one unique MVM.  The
popularity of these two approaches resides in the fact that they rely
upon a broadly available and optimized BLAS level 2 routine:
\pmf{\texttt{gemv}}.  Nevertheless, TVC is far more susceptible to
mode-awareness ---thus indicating that performance is strongly
influenced by the contraction mode $k$--- than a regular second-order
contraction: either vector--matrix ($k=0$) or matrix--vector ($k=1$)
multiplication.  Indeed, in a shared-memory parallel context, it may
not be possible to evenly distribute MVMs among available threads in
the looped TVC algorithm, resulting in suboptimal
performance~\cite{Pawlowski2019}\pmf{: up to 18\% lower performance
  and up to 71\% higher variability across contraction modes.}  For
this reason, authors have recently begun paying attention to
mode-oblivious TVC
procedures~\cite{Pawlowski2019,Pawlowski2020,Martinez2022}.

With the increasing volumes of big data necessary to feed the training
models used in the field of artificial intelligence, it becomes
necessary to provide fast and, more importantly, scalable
implementations of tensor operations~\cite{Shin2021}.
Distributed implementations of TVC and HOPM must execute
efficiently on modern and advanced high performance computing (HPC)
systems composed of clusters of multicore and manycore central
processor units (CPUs), graphics processor units (GPUs), and
high bandwidth memory (HBM) such as those currently found on
pre-exascale systems and future exascale systems.  On this matter,
particular attention must be paid to the streamed memory (i.e., the
total amount of touched memory) and communication performance of such
distributed algorithms, as well as their overall memory utilization
and possible synchronization overheads.

This paper constitutes an extension over our previous article on a
shared-memory, native TVC algorithm for dense
tensors~\cite{Martinez2022}.  In this work, we tackle the parallel
performance of distributed-memory TVC and HOPM algorithms by:
\begin{itemize}
  \item Developing distributed-memory, highly optimized TVC and HOPM
    algorithms and rendering them publicly available in an open-source
    library~\pmf{\cite{Martinez2024}}.
  \item Providing analytical formulae to assess the effect of
    one-dimensional MPI splitting on the streamed memory of TVC and
    HOPM algorithms.
  \item Carrying out exhaustive numerical experiments of distributed
    TVC and HOPM over all contraction and splitting modes using
    different parallel paradigms and three architectures featuring
    CPUs, GPUs and high bandwidth memory.
  \item Assessing the performance of distributed TVC and HOPM against
    state-of-the-art kernels and tensor libraries as well as theoretical
    bandwidth values for single and ad-hoc, mixed precision kernels.
\end{itemize}

The remainder of this paper is organized as follows.
Sections~\ref{sec:related} and \ref{sec:background} are dedicated to
\pmf{giving some related work and describing basic aspects about
  tensors}.  In Section~\ref{sec:algorithm} we detail our distributed
TVC and HOPM algorithms.  We carry out our numerical experiments
\pmf{in} Section~\ref{sec:performance} to assess the performance of
the aforementioned algorithms and, finally,
Section~\ref{sec:conclusions} presents the final conclusions and
future work.

\section{Related work}\label{sec:related}
There are numerous examples of tensor--vector contraction algorithms,
the majority of them being based upon BLAS level 2 and 3
routines~\cite{diNapoli2014,Bassoy2019}.  Consequently, it is not rare
that such algorithms heavily exploit HPC libraries such as Intel
oneMKL, OpenBLAS~\cite{Xianyi2012}, BLIS~\cite{BLIS}\pmf{, as well as
  LIBXSMM~\cite{LIBXSMM}} which provides better performance over
repetitive small matrix--matrix multiplications.  One of the main
challenges of TVC, besides of its memory-bound nature, is its
performance reliance on the contraction mode.  As a result, its global
efficiency can be compromised even with highly optimized BLAS
routines.  Paw{\l}owski et al.\ introduced a sequential,
mode-oblivious TVC algorithm based on \pmf{the} Morton-ordered memory
layout~\cite{Pawlowski2019} and later parallelized it on shared-memory
systems~\cite{Pawlowski2020}.  In this respect, we recently proposed a
shared-memory, mode-oblivious native TVC algorithm for
nonhierarchically stored dense tensors~\cite{Martinez2022} that
naturally distributes the column space of \pmf{the matricized view of
  a tensor} among CPU cores thereby resulting in a specialized,
\emph{nonstandard} BLAS level 2 routine.

With the advent of big data analysis, it has become necessary to
perform tensor calculations on large scale, distributed systems.
Existing tensor libraries typically fall within two main categories.
In the first category, the actual distribution takes place at the
filesystem level via the ``mapReduce'' programming paradigm
popularized by Hadoop, e.g.\ GigaTensor~\cite{Kang2012} and
BIGtensor~\cite{Park2016}, as well as Spark,
e.g.\ CSTF~\cite{Blanco2018}.  In general, these distributed
implementations do not support higher\pmf{-}order tensors and require
additional tensor unfolding operations owing to the explicit parallel
distribution which, in the end, result in filesystem overhead.  The
second category is more interesting from a HPC perspective with
applications making extensive use of the MPI
library~\cite{Walker1994}.
The Algebraic Programming (ALP) framework\footnote{Visit:
\url{https://algebraic-programming.github.io/}} leverages explicit
algebraic annotations to computations for optimization and
auto-parallelization, both shared- and distributed-memory.
ALP/GraphBLAS and the wider GraphBLAS efforts promote the paradigm for
generalized sparse linear algebra~\cite{Yzelman2020,GraphBLAS},
whereas ALP/Dense adds dense linear algebra
support~\cite{Spampinato2023}.
\pmf{Remarkable examples of} tensor frameworks for distributed systems
are CTF~\cite{Solomonik2014} and tiledArray~\cite{Calvin2015}, both
achieving similar levels of parallel performance\pmf{, as well as
  Deinsum/DaCe~\cite{Ziogas2022}}.  CTF aims at minimizing
communication cost, rather than necessarily achieving best
performance.  Its flexibility for multidimensional tensor splitting
implies data redistribution (copying and transposition), hence giving
rise to very low flop-to-byte ratio algorithms that may not achieve
memory bandwidth peak performance.  \pmf{Both CTF and tiledArray}
resort to batched BLAS primitives (e.g.\ from
LAPACK~\cite{Anderson1999}, ScaLAPACK~\cite{Dongarra2011} and others
cited above) with the aforementioned implications on looped TVC
performance.  All things considered, the main advantage offered by
\pmf{these two} frameworks resides in the simplicity and generality of
expressing tensor operations via a domain-specific language (DSL).
\pmf{On the other hand, Deinsum translates Python code into high
  performance binaries, derives data movement-optimal tiling,
  generates corresponding distributed schedules, and can optimize the
  performance of local computations by increasing their arithmetic
  intensity, demonstrating significant speedups over CTF.}

To the best of our knowledge, there are no previous works focusing on
the performance of a mode-oblivious TVC algorithm on distributed
systems using higher\pmf{-}order dense tensors and evaluating the
effects of tensor splitting on bandwidth from both analytical and
numerical perspectives.  Moreover, no other BLAS level 2 TVC algorithm
with oblivious properties has been proposed for hybrid computing, or
compared against state-of-the-art tensor libraries for distributed
systems, or tailored GPU kernels for heterogeneous computing.  In this
regard, the present work showcases the importance of seeking near
theoretical bandwidth performance during distributed TVC operations,
making special emphasis on a real-world tensor operation: the
distributed higher\pmf{-}order power method.  We hope that our
contribution will be used as an example of future development around
HPC libraries oriented towards tensor operations on distributed and
heterogeneous systems.

\section{Background}\label{sec:background}
We briefly introduce some basic concepts about tensors taken mainly
from Kolda \& Bader~\cite{Kolda2009} and reuse the notation introduced
in our previous work~\cite{Martinez2022}.  The font shapes $x$,
$\mathbf{x}$, $\mathbf{X}$ and $\mathcal{X}$ refer to scalars,
vectors, matrices, and tensors, respectively.  We interpret dense
tensors as multidimensional arrays stored in system memory following a
last-order, nonhierarchical storage layout (see
Fig.~\ref{fig:splitting}) which is indeed equivalent to the
multidimensional array ordering used by the C/C++ programming
languages.  Similarly to these languages, we adopt a zero-based
indexing.

By definition, a $d$-order tensor $\mathcal{A} \in \mathbb{R}^{n_0
  \times n_1 \times \ldots \times n_{d-1}}$ is composed of $d$
distinct modes of size $n_i$ and accounts for a total of
$N=\prod_{i=0}^{d-1}n_i$ elements.  The $k$-mode contraction ($0
\leqslant k \leqslant d-1$) of the previous tensor against a given
vector $\mathbf{x} \in \mathbb{R}^{n_k}$ can be written as
$\mathcal{Y} = \mathcal{A} \times_k \mathbf{x}$.  The resulting
tensor, $\mathcal{Y} = \mathbb{R}^{n_0 \times \ldots \times n_{k-1}
  \times n_{k+1} \times \ldots \times n_{d-1}} $, is composed of only
$d-1$ modes: its supposedly $k$-th mode remains of size unity as a
result of the contraction.

For the actual computation of the previous TVC it is practical to
reinterpret the input tensor $\mathcal{A}$ in matrix form.  By
defining $u=\prod_{i=0}^{k-1}n_i$ and $v=\prod_{i=k+1}^{d-1}n_i$ one
can \pmf{build} the matricized form of such tensor $\mathbf{A}^{uv
  \times n_k}$ and its transpose $\mathbf{A}^{n_k \times uv}$.  Now a
looped TVC algorithm can be built upon a series of classical
matrix--vector multiplications over contiguous subsets of
$\mathbf{A}$.  The kind of operation depends on the memory layout and,
for the last-order arrangement adopted herein, one single
matrix--vector multiplication, $\mathcal{Y} = \mathbf{A}^{u \times
  n_k} \times_{k} \mathbf{x}$ is required for the last contraction
mode $k = d-1$; otherwise, $u$ independent, \pmf{equally-sized,}
left-hand sided vector--matrix multiplications \pmf{of the form}
$\mathcal{Y} = \mathbf{x}^\intercal \times_k \mathbf{A}^{n_k \times
  v}$ are necessary for the remaining modes $k < d-1$.  Note that
$u=1$ for $k=0$ and $v=1$ for $k=d-1$.

The \pmf{sequential} looped algorithm described above has been widely
adopted by tensor libraries and can be parallelized at two levels: (i)
the matrix--vector (or vector--matrix) multiplication itself and/or
(ii) the $u$ independent multiplications for $0 < k < d-1$.  HPC
libraries such as Intel \pmf{one}MKL and NVIDIA cuBLAS already provide
optimized (i.e., batched) functions~\cite{Dongarra2016} for subsequent
MVMs: \texttt{cblas\_gemv\_batch\_strided} \pmf{for the former} and
\texttt{cublas\_gemvStridedBatched} \pmf{for the later}.
Nevertheless, all these efforts do not prevent this looped
algorithm\pmf{, either in its sequential or parallel version,} from
being exposed to mode-aware, suboptimal performance and, therefore, it
becomes necessary to find alternative approaches to the looped TVC.

\pmf{The higher\pmf{-}order power method is a generalization of the
  well-known power iteration algorithm applied to matrices and is
  employed to find the best rank-1 approximation of a
  tensor~\cite{Lathauwer2000}.  Given a $d$-order input tensor
  $\mathcal{A}$ and a set of $d$ compatible vectors $\mathbf{x}_0$,
  $\mathbf{x}_1$, $\ldots$, $\mathbf{x}_{d-1}$, the HOPM performs $d$
  external iterations within which $d-1$ TVC operations are carried
  out consecutively omitting, precisely, the contraction along the
  external mode.  This results in a total of $d(d-1)$ tensor
  contractions, which renders the HOPM algorithm an excellent
  benchmark for testing the performance of different TVC
  implementations.  We refer the reader to
  Refs.~\cite{Lathauwer2000,Pawlowski2019} for a canonical
  representation of the HOPM algorithm.

It is worth mentioning that the tensor operations described in this
section refer basically to sequential algorithms.  Shared-memory,
parallel implementations of TVC are discussed by Paw{\l}owski et
al.~\cite{Pawlowski2020} and can also be found in our previous
work~\cite{Martinez2022}.  Distributed-memory implementations are
addressed in the next section.}

\section{\pmf{Distributed algorithms}}\label{sec:algorithm}
This section firstly \pmf{introduces the dTVC} and discusses the
particularities of tensor splitting on distributed systems during
tensor--vector contractions and, secondly, it proposes an optimized
version of the \pmf{dHOPM}.  \pmf{Finally, it comments on the
  data-flow parallelization strategy adopted for the dHOPM based on
  task annotations.}

\subsection{Distributed-memory tensor--vector contraction}\label{sec:dTVC}
To allow for distributed-memory TVC computations, one can split the
given input tensor $\mathcal{A}$ along one or more dimensions.  Such
splitting also influences the output tensor $\mathcal{Y}$ that is
distributed in a similar manner.  The input vector $\mathbf{x}$ that
participates in the contraction can be harmlessly duplicated among the
$p$ distributed processes since, in general, we can assume that $uv
\gg n_k$.  In this work, we only consider one-dimensional cuts of both
the input and output tensors for three main reasons: (i) it yields
minimum communication~\cite{Pawlowski2020}, (ii) it gives best
computational performance as it does not incur additional tensor
unfolding operations, and (iii) it greatly simplifies tensor
distribution and reassembly.  The major disadvantage of 1D splitting
resides in the fact that the maximum number of processes that can be
used is determined by $\max(n_i)$.  Such processes can be bound to
\pmf{CPU cores,} NUMA nodes, GPU devices or even entire compute nodes
thereby maximizing the potential for parallelism.

\begin{figure}[tbp]
  \centering
  \includegraphics[width=0.45\textwidth]{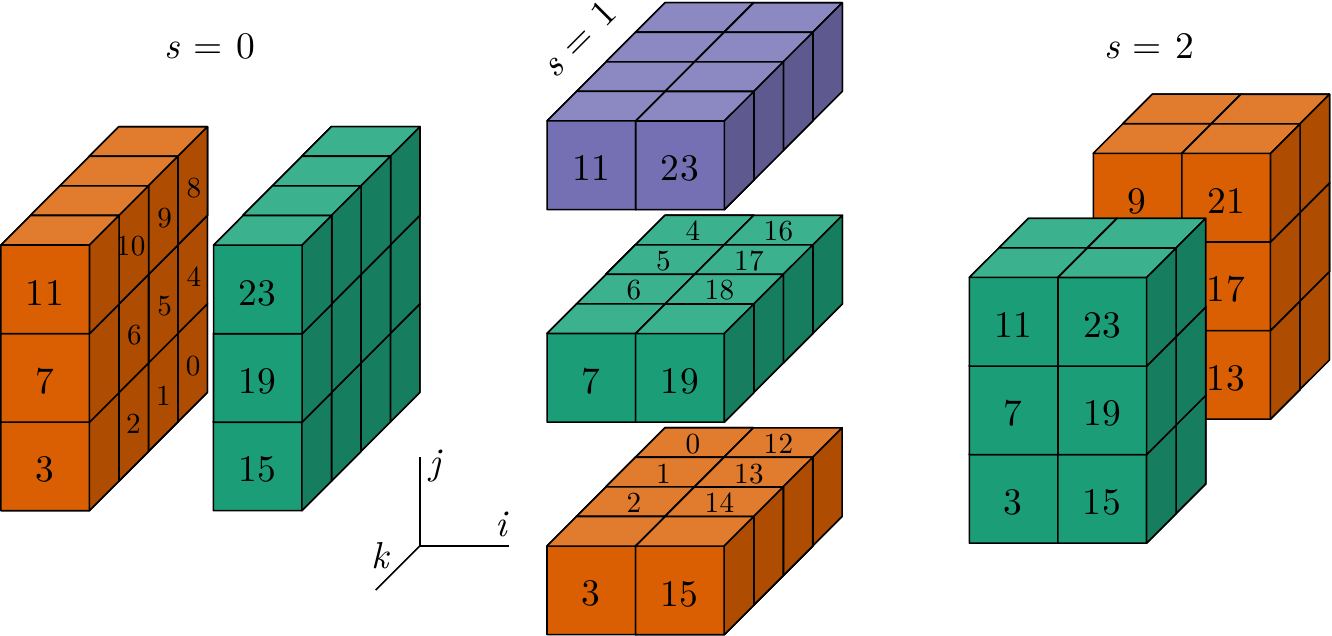}
  \caption{One-dimensional optimal splittings ($s=0,1,2$) using up to
    three processes ($p=3$) \pmf{for} a third-order tensor
    $\mathcal{A}^{2 \times 3 \times 4}$ stored in last-order,
    nonhierarchical memory layout.}
  \label{fig:splitting}
\end{figure}

Let us define the contraction mode $k$ and splitting dimension $s$.
Since we store tensors \pmf{in} last-order layout, a splitting along
the first dimension of $\mathcal{A}$ guarantees the contiguity between
elements of the resulting $\mathcal{A}^{(p)}$ distributed subtensors:
$\mathcal{A} = \bigsqcup_p \mathcal{A}^{(p)}$ with $\mathcal{A}^{(p)}
\in \mathbb{R}^{[n_0/p] \times n_1 \times \ldots \times n_{d-1}}$.  In
other words, the distributed tensors can be reassembled into a global
tensor straightforwardly as shown in Fig.~\ref{fig:splitting} with
$s=0$.  The symbol $\bigsqcup_{\pmf{p}}$ indicates the disjoint union
of \pmf{$p$} tensors and \pmf{the brackets in} $[n_0/p]$ refer to the
\emph{optimal} division between $n_0$ and $p$.  Usually, it
corresponds to the ceiling division \pmf{$\lceil n_0/p \rceil$}
although, for \emph{vectorization} purposes, we employ a heuristic
technique that seeks to promote quotients that are multiple of the
vector length (e.g., 512~bits).  This may override (i.e., lower) the
value of $p$ for an optimal splitting; for instance,
Fig.~\ref{fig:splitting} with $s=2$ results in $[4/3] \rightarrow
4/2$, that is \pmf{$p \rightarrow 2$, so that} only two of the three
requested processes are actually employed.  Following the same
reasoning, a cut along the last dimension, $[n_{d-1}/p]$, results in
the most indirect tensor distribution and later assembly: in the
particular case of a second-order tensor, this corresponds to
splitting along columns \pmf{($s=1$)} a matrix stored in row-major
memory layout.

We now proceed with the tensor--vector contraction algorithm in terms
of both intra- and inter-process data movement.  A general\pmf{,
  BLAS-like expression for the \pmf{distributed} TVC operation with
  typical scalars $\alpha$ and $\beta$} can be written \pmf{for $k
  \neq s$} as
\begin{equation}\label{eq:TVC-k!=s}
  \mathcal{Y} := \alpha \bigsqcup_p \mathcal{A}^{(p)} \times_{k \neq s} \mathbf{x} + \beta \bigsqcup_p \mathcal{Y}^{(p)},
\end{equation}
otherwise
\begin{equation}\label{eq:TVC-k==s}
  \mathcal{Y} := \alpha \sum_p \mathcal{A}^{(p)} \times_{k = s} \mathbf{x}^{(p)} + \beta \bigsqcup_p \mathcal{Y}^{(p)},
\end{equation}
when the contraction mode and the splitting dimension coincide
($k=s$).  This second case remains suboptimal because each distributed
TVC operation yields $p$ subtensors with the exact same size
\pmf{$N/n_k$} as the global result, which translates into more local
computations and hence streamed memory compared to the first case.
Additionally, Eq.~\eqref{eq:TVC-k==s} demands a collective summation
of all these equally-sized subtensors, element by element, to get the
global tensor, hence the sum symbol.  In contrast,
Eq.~\eqref{eq:TVC-k!=s} incurs much less communication because each
subtensor only consists of \pmf{approximately} $N[n_s/p]/(n_kn_s)
\approx N/(n_kp)$ elements.  Since $k=s$ leads to further
communication as well as computation, one should avoid contracting
tensors along their splitting dimension as much as possible.  What is
more important, it is strongly encouraged to work with the distributed
output tensors and only construct the global tensor when strictly
needed because communication is slow and available memory may also be
an issue.  Finally, $\mathbf{x}^{(p)}$ in Eq.~\eqref{eq:TVC-k==s}
indicates that the contraction operates on a disjoint subset of
$\mathbf{x}$ instead of the entire vector.  Getting back to the
previous matrix analogy, Eq.~\eqref{eq:TVC-k==s} is identical to a
series of \pmf{matrix--vector} multiplications applied to subsets of
both $\mathbf{x}$ and $\mathbf{A}$, where the matrix is distributed
along \pmf{columns ($k=s=1$) or a series of vector--matrix
  multiplications when the matrix is distributed along rows
  ($k=s=0$).}

An important aspect related to the global construction of the output
tensor is the contiguity of the elements being assembled.  As
\pmf{previously shown in Fig.~\ref{fig:splitting}}, using \pmf{the}
  last-order layout with $s=0$ simply requires to contiguously gather
  all rows from each process in order to build the global tensor.  On
  the other hand, for $s>0$ it becomes necessary to communicate
  fragments of data from each distributed tensor and interleave them
  inside the global tensor.  To this end, one can resort to the
  matricized views of $\mathcal{Y}^{(p)}$ and $\mathcal{Y}$, both
  split along the $s-1$ dimension, to determine the groups of columns
  to be collected from each process.  This results in
  $w=\prod_{i=0}^{s-2}n_i$ communication messages per process compared
  to a single one for $s=0$ (likewise, $w=1$ for $s=1$) for gathering
  a total of $N/n_k$ elements.  Therefore, increasing values of $s$
  are expected to negatively impact parallel performance.  However,
  this adversity can be overcome by trading messages for further data
  movement: firstly, local memory requirements are duplicated to hold
  two global tensors; secondly, one collective per process is employed
  to gather all the distributed tensors consecutively in memory inside
  the first global (disjoint) buffer; and, lastly, columns are copied
  in groups to the second buffer in order to compose the global
  (joint) tensor thereby incurring in the additional movement of
  $N/n_k$ elements within local memory.

\subsection{Distributed-memory higher-order power method}\label{sec:dHOPM}

In the next paragraphs we analyse the HOPM in terms of intra-process
data movement exclusively.  In the particular case of hypersquare
tensors ($n_i = n$), a sequential implementation of the HOPM that
can be found in Refs.~\cite{Lathauwer2000,Pawlowski2019} incurs a
certain amount of touched memory
\begin{equation}\label{eq:mseq}
  m_{\rm seq} = n^d + 2\sum^{d-1}_{k=2}{n^k} + (d + 3) n ,
\end{equation}
for each one of the aforementioned $d$ external iterations.  The first
term at the r.h.s.\ of Eq.~\eqref{eq:mseq} corresponds to the size of
the input tensor and the second term represents the streamed memory
for intermediate tensors.  The third term refers to the touched memory
of the final output tensor (a vector of size $n$), all the
participating input vectors $(d-1)n$, and the normalization step $3n$.
Finally, the total amount of streamed memory of the HOPM algorithm
accounted for the $d$ external iterations is simply $M_{\rm seq} = d
m_{\rm seq}$.

In contrast to Eq.~\eqref{eq:mseq}, the touched memory expression of
the distributed HOPM algorithm loses the symmetry of its sequential
counterpart.  Each parallel process incurs $M_{\rm par} = s m_{{\rm
    par}, j<s} + m_{{\rm par}, j=s} + (d-1-s) m_{{\rm par}, j>s}$,
with
\begin{align}\label{eq:mpar}
  m_{{\rm par}, j=s} &= \left[ \frac{n^d}{p} \right] + 2 \sum^{d-1}_{k=2} \left[ \frac{n^k}{p} \right]
                  + 4 \left[ \frac{n}{p} \right] + (d-1)n \nonumber \\
                  &\approx \frac{m_{\rm seq}}{p} + \frac{p-1}{p}(d-1)n ,
\end{align}
\pmf{when the external iteration $j$ coincides with the splitting
  dimension $s$}.  Square brackets denote the optimal division but,
for the sake of simplicity, we assume a regular division thereby
approximating the previous equation so that it can be related to
Eq.~\eqref{eq:mseq}.  In fact, the second term of Eq.~\eqref{eq:mpar}
reveals that the duplication of the input vectors on each MPI process
causes a memory overhead over the sequential algorithm.  For the
remaining $d-1$ external iterations, we get the following expression
\begin{align}\label{eq:mpar2}
m_{{\rm par}, j \neq s} &\approx \frac{m_{\rm seq}}{p} + \frac{p-1}{p} \left(2\sum^{d-s-l}_{k=2}{n^k} + (d+2)n \right) \nonumber \\
                    &= m_{{\rm par}, j=s} + \frac{p-1}{p} \left(2\sum^{d-s-l}_{k=2}{n^k} + 3n \right) ,
\end{align}
with $l=0$ if $j<s$, otherwise $l=1$ if $j>s$.  It can be noticed that
Eq.~\eqref{eq:mpar2} incurs further overhead over Eq.~\eqref{eq:mpar}
because, contrary to the previous case, now the contraction mode and
the splitting index \pmf{will} coincide ($k=s$) \pmf{during} one of
the $d-1$ inner TVCs thus yielding larger partial subtensors as
discussed in Section~\ref{sec:dTVC}.  This is indeed a very
interesting aspect of the dHOPM algorithm: it will lead to $d-1$
underperforming contractions.  The final expression for the dHOPM
touched memory per process can be written as
\begin{align}\label{eq:Mpar}
\begin{aligned}
  M_{\rm par} &\approx \overbrace{ \frac{M_{\rm seq}}{p} + \frac{p-1}{p} (d-1)(d+3)n }^{M_{\rm par,min}} \\
  & \phantom{{}\approx \frac{M_{\rm seq}}{p}{}} + \frac{p-1}{p} \left(s \sum^{d-s}_{k=2}{2n^k} + (d-s-1) \sum^{d-s-1}_{k=2}{2n^k} \right) ,
\end{aligned}
\end{align}
which reveals that there is a constant, minimum amount of streamed
memory $M_{\rm par,min}$ caused by the duplicates of the input vectors
and the subpar contractions of Eq.~\eqref{eq:mpar2}.  On the other
hand, the last term of Eq.~\eqref{eq:Mpar} is highest for $s=0$ and
cancels out for $s=d-1$.  It is easy to derive a recursive form
\begin{align}\label{eq:Mpar2}
\begin{aligned}
  M_{{\rm par}, s-1} &= M_{{\rm par}, s} \\
  &+ \frac{p-1}{p} \left( (d-s-1)2n^{d-s} + (s-1)2n^{d-s+1} \right) ,
\end{aligned}
\end{align}
that elucidates a linear increase of data movement with the number of
processes and a more complex, nonlinear relationship with the
splitting dimension.  \pmf{Contrary to the dTVC algorithm and its
  later assembly process described at the end of
  Section~\ref{sec:dTVC}, for the dHOPM} one should avoid splitting
tensors along their first dimensions, especially when using many
processes.  \pmf{This will significantly reduce the movement of data
  and yield faster computations.}

\begin{figure}[tbh]
  \centering
  \subfloat[]{
    \includegraphics[width=0.4\textwidth]{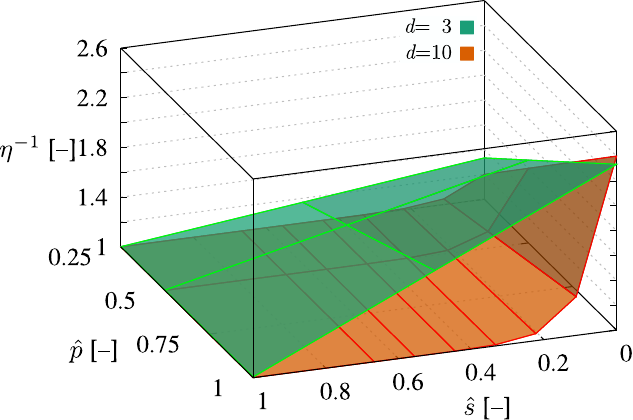}
  }
\\
  \subfloat[]{
    \includegraphics[width=0.4\textwidth]{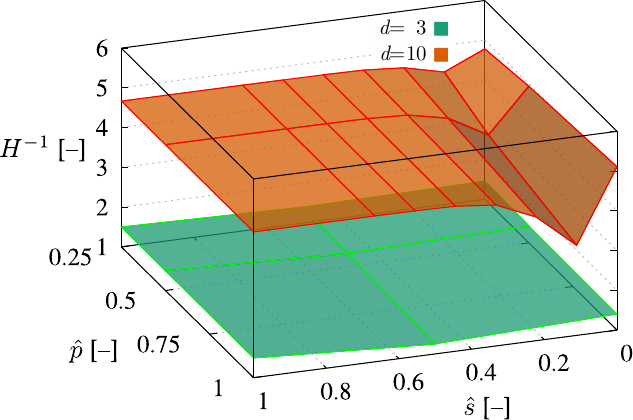}
  }
  \caption{(a) memory ratio ($\eta^{-1}$) incurred by the classical,
    dHOPM implementation and (b) memory ratio ($H^{-1}$) between this
    canonical version and dHOPM$_3$ as a function of the
    nondimensional number of processes ($\hat{p}$) and the
    nondimensional splitting dimension ($\hat{s}$).  Results
    correspond to the third- and tenth-order tensors of
    Table~\ref{tab:tensors}.}
  \label{fig:touchMem}
\end{figure}

Let us consider the following nondimensional variables: $\hat{p} =
p/n$, $\hat{s} = s/(d-1)$, and $\eta^{-1}(\hat{p},\hat{s}) = p M_{\rm
  par}/M_{\rm seq}$.  Figure~\ref{fig:touchMem}(a) shows the memory
ratio $\eta^{-1}$ of the classical dHOPM for the third- and
tenth-order hypersquare tensors considered in this work (see
Table~\ref{tab:tensors}).  In both cases, the data movement more than
doubles for $\hat{s} = 0$ and $\hat{p} = 1$.  With $d=3$ the memory
ratio increases steadily for decreasing values of $\hat{s}$ while, for
$d=10$, memory effects are limited to a narrow region of low values of
$\hat{s}$.  As expected, the touched memory tends to augment with the
number of processes and, in any case, if the tensor is split along its
last dimension, then $M_{\rm par} \approx M_{\rm par,min} \approx
M_{\rm seq}/p$.  A similar trend is observed for the other tensors of
Table~\ref{tab:tensors} which are not shown in
Fig.~\ref{fig:touchMem}(a) for the sake of conciseness.  The only
exception to this rule comes from the second-order tensor (matrix)
where $M_{\rm par, min}$ only varies with $p$, as it can be deduced
from Eq.~\eqref{eq:Mpar}, and its value can be significantly larger
than those associated with higher\pmf{-}order tensors.  Finally, it is
worth noting that the expressions \eqref{eq:mseq}--\eqref{eq:Mpar2}
and Fig.~\ref{fig:touchMem} are still representative of nonhypersquare
tensors provided that the size assigned to each dimension remains
about the same order of magnitude.

\begin{algorithm}[tbh]
\caption{Distributed-memory, three-buffer higher\pmf{-}order power method algorithm (dHOPM$_3$).}\label{alg:HOPM}
\small
\DontPrintSemicolon
\SetKwInOut{myIn}{Input}
\SetKwInOut{myInOut}{Input/Output}
\SetKwInOut{buffers}{Buffers}
\myIn{distributed $d$-order tensor $\mathcal{A}^{(p)}$.}
\myInOut{$d$ global vectors $\mathbf{x}_0$, $\mathbf{x}_1$, $\ldots$, $\mathbf{x}_{d-1}$.}
\buffers{inputs $\mathcal{Y}_{k-1}^{(p)}$, $\mathcal{W}_{j-1}^{(p)}$; output $\mathcal{Y}_{k}^{(p)}$.}
\BlankLine
\PrintSemicolon
\For{$i \leftarrow 0, 1, \ldots$}{
  \For(\tcp*[f]{Ext.~iteration}){$j \leftarrow 0$ \KwTo $d-1$}{
    $\pmf{\lambda} \leftarrow 0$  {\bf if} $j>0$, {\bf else} $\pmf{\lambda} \leftarrow 1$ \;
    $\pmf{\mu}  \leftarrow \max(\pmf{\lambda}, j)$ \;
    $\pmf{\nu} \leftarrow d-1$  {\bf if} $j<d-1$, {\bf else} $\pmf{\nu} \leftarrow d-2$ \;
    \If(\tcp*[f]{TVC(1/3)}){$j < 2$}
    {
      $\mathcal{Y}_{\pmf{\lambda}}^{(p)} := \mathcal{W}_j^{(p)} \leftarrow \mathcal{A}^{(p)} \times_{\pmf{\lambda}} \mathbf{x}_{\pmf{\lambda}}$ \;
    }
    \Else
    {
      $\mathcal{Y}_{\pmf{\mu}-1}^{(p)} := \mathcal{W}_j^{(p)} \leftarrow \mathcal{W}_{j-1}^{(p)} \times_{\pmf{\mu}-1} \mathbf{x}_{\pmf{\mu}-1}$ \;
    }
    $\mathcal{Y}_{\pmf{\mu}+1}^{(p)} \leftarrow \mathcal{Y}_{\pmf{\mu}+\pmf{\lambda}-1}^{(p)} \times_{\pmf{\mu}+1} \mathbf{x}_{\pmf{\mu}+1}$ \tcp*[r]{TVC(2/3)}
    \For(\tcp*[f]{TVC(3/3)}){$k \leftarrow \pmf{\mu}+2$ \KwTo $\pmf{\nu}$}{
      $\mathcal{Y}_{k}^{(p)} \leftarrow \mathcal{Y}_{k-1}^{(p)} \times_k \mathbf{x}_k$ \;
    }
    \If{$j \neq s$}
    {
      $\mathbf{x}_j \leftarrow \sum_p \mathcal{Y}_{\pmf{\nu}\phantom{k}}^{(p)}$ \tcp*[r]{Array reduction}
    }
    \Else
    {
      $\mathbf{x}_j \leftarrow \bigsqcup_p \mathcal{Y}_{\pmf{\nu}\phantom{k}}^{(p)}$ \tcp*[r]{Array gather}
    }
      $\mathbf{x}_j \leftarrow \mathbf{x}_j /||\mathbf{x}_j||$ \;
  }
}
\end{algorithm}

The classical dHOPM \pmf{analyzed above} can be further optimized for
distributed systems as shown in Algorithm~\ref{alg:HOPM}.  This novel
approach, which makes use of three buffers \pmf{(dHOPM$_3$), employs}
two \pmf{of them} $\mathcal{Y}_{k-1}$ and $\mathcal{Y}_{k}$ to save
intermediate tensors and a third \pmf{one} $\mathcal{W}$ to hold
previously computed data in memory ($k=\pmf{\lambda}$ or
$k=\pmf{\mu}-1$, see lines 7 or 9) for the next iteration when $j \geq
2$ (line 9).  This allows us to save a total of $(d-1)(d-2)/2$
contractions and, as the order of the original \pmf{input} tensor
keeps increasing, dHOPM$_3$ can save up to half of the total
contractions required by a canonical, two-buffer algorithm such as the
one from Paw{\l}owski et al.~\cite{Pawlowski2019}.  What is more
important, the skipped contractions are precisely the most
computationally expensive ones since they are carried out on the
largest input tensors, which results in even better computational
speedups.  Another aspect of Algorithm~\ref{alg:HOPM} worth mentioning
is that it inherently exploits the distributive property of
consecutive TVCs and, consequently, the summation required by
Eq.~\eqref{eq:TVC-k==s} when \pmf{$k = s$ (which implies $j \neq s$)}
is delayed until the normalization step, thereby reducing the
communication to only $n_j$ elements per process at line 14; in other
words, the consecutive TVC operations at lines 6--12 are done
exclusively on partial, distributed subtensors without incurring in
global assembly penalizations.  Alternatively, the case $j=s$ implies
the disjoint union of $\mathbf{x}_j$ (line 16) where each process
exchanges their calculated portion $\mathbf{x}_j^{(p)}$ for an
aggregated communication of $n_j$ elements per process.  Finally, the
resulting vector is normalized locally by each process at line 17.

Now we can define $H^{-1} (\hat{p},\hat{s})$ as the streamed memory
ratio between the classical dHOPM and our optimal implementation
dHOPM$_3$.  Figure~\ref{fig:touchMem}(b) shows the evolution of
$H^{-1}$ for the previously analyzed third- and tenth-order tensors.
We obtain two approximately flat surfaces, especially for the
low\pmf{-}order tensor, which indicates that this ratio is almost
independent of $\hat{p}$ and $\hat{s}$.  Our optimized algorithm
economizes about 1.5$\times$ of the touched memory for $d=3$ and
roughly a fivefold for $d=10$ (with the presence of a minimum of about
3.3$\times$).  All things considered, Figs.~\ref{fig:touchMem}(a)--(b)
demonstrate that dHOPM$_3$ can save up to one order of magnitude of
streamed memory for higher\pmf{-}order tensors in comparison to a
canonical HOPM implementation with naively split tensors.

As previously mentioned, Eqs.~\eqref{eq:mpar}--\eqref{eq:Mpar2} do not
account for inter-process data movement.  On the one hand, for all the
external iterations except one, the resulting vector $\mathbf{x}_j$
must be entirely reduced by all processes (see line 14 of
Algorithm~\ref{alg:HOPM}), otherwise only a portion of this array is
gathered (line 16).  The actual amount of data being both transferred
and computed ultimately depends on the algorithm employed underneath
the functions \texttt{MPI\_Allgather} and \texttt{MPI\_Allreduce} and
it is likely to change at runtime based on the message size, the
number of processes and the network topology~\cite{Chan2007}.  For
example, the straightforward \pmf{r}ing algorithm, which is
bandwidth-optimal, suitable for large messages, and works with any
value of $p$, yields $4n(p-1)/p$ extra touched memory per process, a
quantity that can be appended to the term $M_{\rm par,min}$ in
Eq.~\eqref{eq:Mpar}.  It can be easily inferred that the inter-process
contribution reaches its maximum for $d=2$ and $\hat{p}=p/n=1$,
increasing the value of $M_{\rm par,min}$ by up to $4/7 \approx 57\%$.
In the case of hypersquare tensors, this constant number remains
independent of the splitting dimension and can be halved by storing
separate copies of the input tensor $\mathcal{A}$ with different
splittings in order to foster partial gatherings of $\mathbf{x}_j$.

\begin{figure*}[tbh]
  \centering
  \subfloat[Full dHOPM$_3$ execution ($i=0$)]{\includegraphics[width=0.9\textwidth]{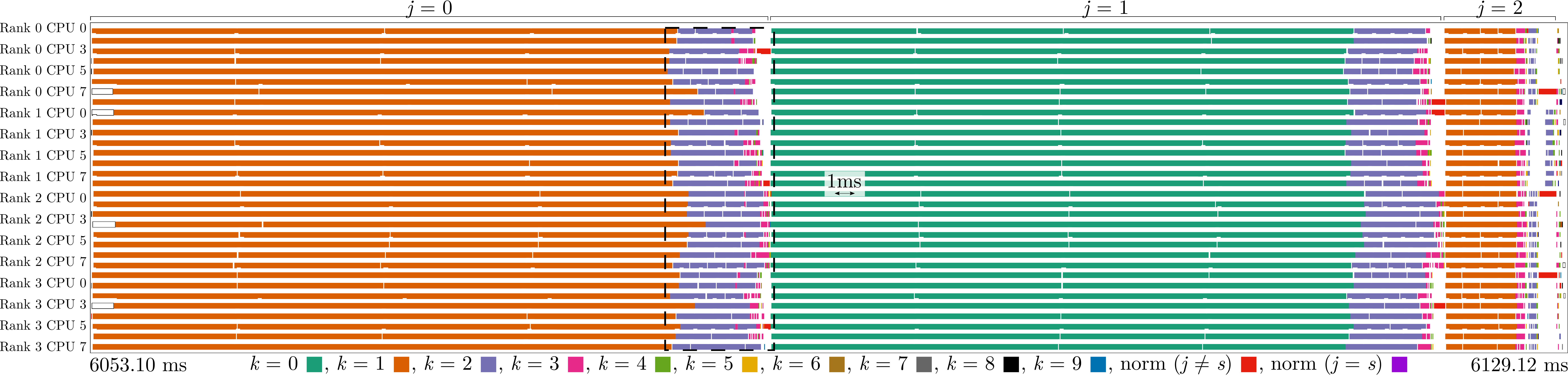}}
\\
  \subfloat[Zoom-in on the last $k$ internal contractions of iteration $j=0$]{\includegraphics[width=0.9\textwidth]{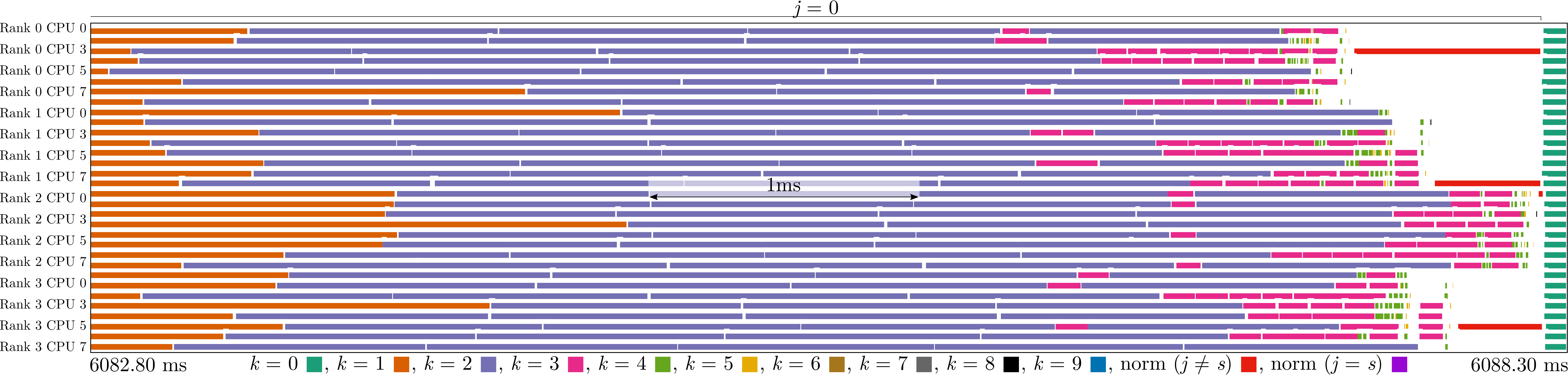}}
  \caption{Paraver traces from the hybrid execution of (a) the
    dHOPM$_3$ algorithm and (b) a zoom-in view of the last $k$
    internal contractions of iteration $j=0$ using the tenth-order
    tensor of Table~\ref{tab:tensors} split along its last dimension.
    Traces obtained on MN4 using 4 MPI processes, 8 cores per process,
    and 32 OmpSs-2 tasks per process for illustration purposes.}
  \label{fig:HOPM-traces}
\end{figure*}
\subsection{\pmf{Task-based parallelization of the dHOPM$_3$ algorithm}}\label{sec:dHOPM-tasks}
We take the opportunity to describe herein the shared-memory parallel
strategy adopted for dHOPM$_3$ in which every contraction (lines 7, 9,
10\pmf{,} and 12 of Algorithm~\ref{alg:HOPM}) takes advantage of
task-based parallelization.  In contrast to ubiquitous OpenMP parallel
for loops, tasks permit to overlap computations corresponding to
different TVC instances, thereby maximizing the parallel performance
over fork-join.  This key novel parallel design is \pmf{exemplified
  below for a single TVC operation, defined by the \texttt{getvc}
  function~\cite{Martinez2024}, which is split in several tasks:}
{\small
  \begin{alltt}
  \teal{if} (trans \gray{==} iZero) \{ \textit{\gray{// Matrix-vector (row-major)}}
    ...
    \textbf{\teal{for}} (\textbf{\purple{intT}} i \gray{=} iZero; i \gray{<} m; i \gray{+=} bsM) \{
      ...
      \textit{\brown{#pragma omp task depend(in : x[iZero:     n]) \textbackslash
                       depend(in : a[i*lda:bsMA*n]) \textbackslash
                       depend(out: y[i    :bsMA  ])}}
      getvc(layout, trans, bsMA, n,
            alpha, a \gray{+} i\gray{*}lda, lda,
            x, incx, y, beta, y \gray{+} i, incy);
    \}
  \} \teal{else} \{ \textit{\gray{// Vector-matrix (row-major)}}
    ...
    \textbf{\teal{for}} (\textbf{\purple{intT}} j \gray{=} iZero; j \gray{<} n; j \gray{+=} bsN) \{
      ...
      \textit{\brown{#pragma omp task depend(in : x[iZero :m   ]) \textbackslash
                       depend(in : a[startl:size]) \textbackslash
                       depend(out: y[j     :bsNA])}}
      getvc(layout, trans, m, bsNA,
            alpha, a \gray{+} startl, lda,
            x, incx, y, beta, y \gray{+} j, incy);
    \}
  \}
\end{alltt}
}

\pmf{Each TVC is assigned a certain number of tasks (i.e.,
  granularity) inside a loop that can be executed concurrently.  We
  establish task data dependencies via \texttt{depend} clauses for the
  input vector \texttt{x} and tensor \texttt{a} as well as the output
  tensor \texttt{y} to ensure the correct order of execution of
  different TVCs and normalization steps as indicated in
  Algorithm~\ref{alg:HOPM}.  The task granularity is such that there
  are more tasks than threads, which permits the overlapping of
  subsequent TVC operations.  This is clearly seen in
  Fig.~\ref{fig:HOPM-traces},} which corresponds to a real execution
carried out on MN4, instrumented with Extrae and visualized with
Paraver\footnote{Extrae and Paraver are available at:
\url{https://tools.bsc.es/}} for a tenth-order tensor distributed
among four processes.  While Fig.~\ref{fig:HOPM-traces}(a) covers a
full dHOPM$_3$ execution ($i=0$), the figure below is a zoomed version
towards the end of the first external iteration $j=0$.  \pmf{In both
  figures,} subsequent TVC operations \pmf{are} represented with
rectangles \pmf{(tasks)} of different colors.  \pmf{As previously
  mentioned, threads can start executing blocks of different colors as
  quickly as possible.}  There are still some relatively small white
gaps that evidence code sections with no CPU usage and correspond
exclusively to the final normalization phase (lines 13--1\pmf{6} of
Algorithm~\ref{alg:HOPM}) where synchronous collective communication
takes place and MPI processes must wait for the slowest one due to
system imbalance.  Both subfigures also reveal how computational costs
greatly reduce after each contraction as the resulting tensor loses
one dimension and its size decreases by about one order of magnitude.

\pmf{Contrary to tasks, a shared-memory strategy based upon the
  fork-join paradigm (see the code snippet from
  Ref.~\cite{Martinez2022}) and, by extension a CUDA-based
  implementation, presents one implicit synchronization per
  contraction.  Consequently, a Paraver trace of such parallel
  execution (not shown here for the sake of conciseness) would have
  shown white gaps towards the end of each TVC where idle threads must
  wait to the slowest one to finish its corresponding computation,
  close the parallelism (join) and, finally, reopen it for the next
  contraction (fork).  All these additional periods of CPU inactivity
  increment the computational time and the application imbalance that
  ultimately result in an overall lower speedup with respect to the
  task-based implementation.}

\section{Performance evaluation}\label{sec:performance}
This section evaluates the performance of the distributed TVC and HOPM
algorithms detailed in previous sections of this manuscript after
describing the numerical environment setup.  The complete source code
of \pmf{our} dTVC library \pmf{is} made publicly available under
the GPLv3 license \pmf{in Ref.~\cite{Martinez2024}}.

\subsection{HPC systems and code setup}
We have targeted three different hardware architectures available at
the Barcelona Supercomputing Center (BSC).  The first one is the
general-purpose MareNostrum 4 (MN4) supercomputer, whose compute nodes
are composed of two Intel Xeon Platinum 8160 CPUs with 24 cores each
(hyperthreading disabled), 33~MiB of L3 cache, and AVX-512 SIMD
instructions.  The system memory is based on DDR4 and peaks at a
theoretical bandwidth of 128~GB/s per socket (256~GB/s per node).  The
second architecture is the CTE-ARM cluster where each compute node
integrates four Fujitsu ARM A64FX CPUs with 12 cores, a last level
cache of 8~MiB, and 512-bit SVE instructions.  Each socket has 8~GiB
of installed HBM2 reaching theoretically 256~GB/s (1024~GB/s per
node).  Lastly, the third architecture is the CTE-POWER cluster
equipped with four NVIDIA Volta V100 GPUs per compute node.  Each
graphic card features 16~GiB of HBM2 reaching a theoretical peak
bandwidth of 900~GB/s (3600~GB/s per node).

\begin{table}[tbh]
\centering
  \caption{Number of 8-byte floating-point elements and corresponding
    memory footprint (measured in GB) for the hypersquare tensors used
    in this work.}
  {\small
  \begin{tabular}{crc}
    \toprule
    Order & Elements & Memory \\
    \midrule
    \phantom{1}2 & $30623^{2\phantom{0}}$ & 7.50 \\
    \phantom{1}3 & $979^{3\phantom{0}}$   & 7.51 \\
    \phantom{1}4 & $175^{4\phantom{0}}$   & 7.50 \\
    \phantom{1}5 & $63^{5\phantom{0}}$    & 7.94 \\
    \phantom{1}6 & $31^{6\phantom{0}}$    & 7.10 \\
    \phantom{1}7 & $19^{7\phantom{0}}$    & 7.15 \\
    \phantom{1}8 & $13^{8\phantom{0}}$    & 6.53 \\
    \phantom{1}9 & $10^{9\phantom{0}}$    & 8.00 \\
              10 & $8^{10}$   & 8.59 \\
    \midrule
    Avg.        & $9.4 \times 10^9$ & 7.54 \\
    \bottomrule
  \end{tabular}
  }
  \label{tab:tensors}
\end{table}

The tensors used in this work are listed in Table~\ref{tab:tensors}.
We choose hypersquare and up to tenth-order dense tensors.  In
principle, they are filled with 8-byte floating-point numbers (i.e.,
doubles) following other works~\cite{Pawlowski2019,Martinez2022}.
Lower\pmf{-}precision floats will be introduced later in
Section~\ref{sec:mixed-precision}.  When possible, each tensor
dimension size has been intentionally chosen to prevent it from being
a multiple of the vector length ---which is 8 when using doubles and
the 512-bit SIMD/SVE instruction set--- with the aim of fostering
scenarios showcasing unaligned memory accesses as well as peeling and
remainder loops.  The aforementioned tensors average 7.5~GB of memory
in order to guarantee that TVC operations are limited by the system
memory bandwidth while fitting within the rather small HBM2 memory
incorporated in two of the three systems described above.
Consequently, larger tensors with vector-friendly dimension sizes are
expected to yield better performance figures.

With regard to code compilation, we utilize the LLVM infrastructure
unless otherwise stated.  In particular, we employ the Clang++ 16.0
compiler that supports both the OpenMP~\cite{Dagum1998} and
OmpSs-2~\cite{Perez2017} programming languages allowing for data-flow
parallelization via task \pmf{annotations}.  In order to obtain the
maximum performance, we use the flag \texttt{-march= native} to enable
CPU specific optimizations.  The combination of flags \texttt{-Ofast}
and \texttt{-mprefer-vector-width=512} are used to exploit 512-bit
vector instructions.  In addition to this, the flag \texttt{-fopenmp}
enables OpenMP pragma directives used for both vectorization and
parallelization of the code: unless otherwise stated, shared-memory
parallelization is achieved via fork-join, ``parallel for'' loops.
Furthermore, the flag \texttt{-flto} enables interprocedural (i.e.,
link-time) optimizations.  Other code parameters are the alignment of
buffers to transparent huge pages of 2~MiB and the loop unrolling
factor set to 8.  For GPU devices, we make use of the latest cuBLAS
10.2 library officially supported by the architecture vendor.

Each benchmark steadily executes a particular TVC or HOPM computation
during five seconds allowing tens or even hundreds of kernel calls to
retrieve statistical figures.  Strong scalability tests are conducted
on up to 128 MPI processes which are bound to distinct NUMA nodes or
GPU devices to allow for hybrid or heterogeneous parallelization,
respectively.  We employ the MPI libraries provided by the hardware
manufacturers: Intel MPI on MN4, Fujitsu MPI on CTE-ARM and IBM
Spectrum on CTE-POWER.  Besides, we choose NVIDIA NCCL over CUDA-aware
MPI functions for best performance.  We employ three performance
metrics: (i) the normalized bandwidth that is obtained after dividing
by the theoretical peak bandwidth corresponding to each architecture,
(ii) the kernel throughput measured in iterations per second (it/s)
and (iii) the normalized kernel throughput measured in iterations per
second per MPI process (it/sp).  We remind the reader that the memory
bandwidth is the touched (i.e., streamed) memory per unit of time and
the STREAM benchmark~\cite{McCalpin1995}, particularly the
\texttt{triad} function, reports the following figures on each NUMA
node or GPU device: 104.5~GB/s (81.6\% of the theoretical peak) on
MN4, 153.7~GB/s (60\%) on CTE-ARM, and 556.2~GB/s (61.8\%) on
CTE-POWER.

\begin{figure*}[htb]
  \centering
  \martinezFigLabel{0.9\textwidth}{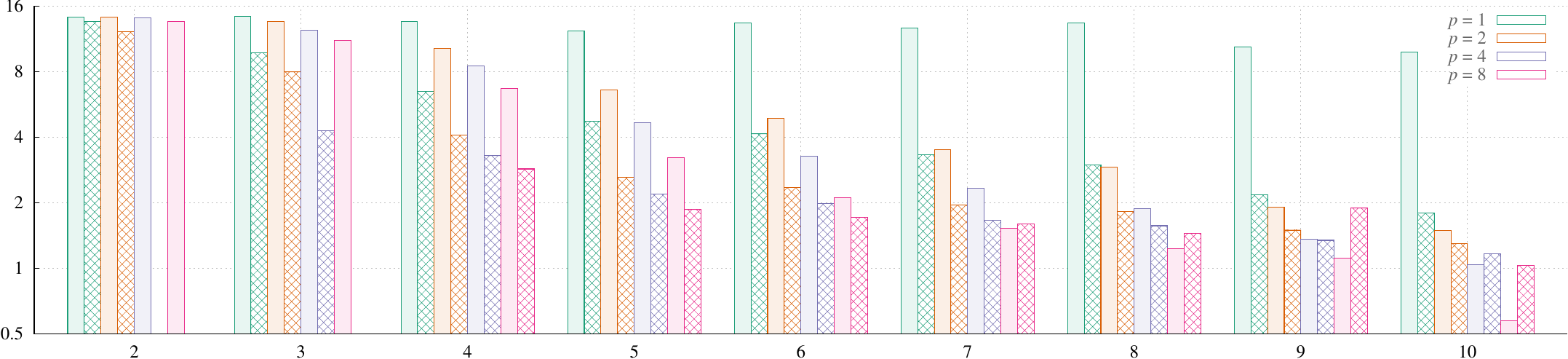}
  {Tensor order}{[--]}
  {Norm. throughput}{[it/sp]}
  \caption{dTVC normalized throughput (including global assembly of
    the output tensor) measured as the number of iterations per second
    per MPI process (it/sp) over all contraction modes and all
    splitting dimensions for the tensors of Table~\ref{tab:tensors}.
    Strong scalability results using up to 8 MPI processes on MN4
    employing our proposed distributed TVC algorithm (solid fill) and
    the CTF implementation (pattern fill).}
  \label{fig:HOPM-CTF}
\end{figure*}

\subsection{State-of-the-art performance of the tensor--vector contraction}
We find it convenient to give the reader a notion about the expected
parallel performance of the dTVC algorithm using the software
available in the literature.  To this end, we provide a comparison of
our proposed solution against the massively parallel tensor
contraction \pmf{(CTF) framework~\cite{Solomonik2014}
  (Deinsum~\cite{Ziogas2022} does not currently support tensor--vector
  contractions)}.  This framework enables hybrid parallelism via MPI
and MKL libraries and also makes extensive use of the optimized
kernels provided by the later, namely: BLAS, LAPACK, and ScaLAPACK.
We compiled CTF against the latest Intel software suite available on
MN4 (Intel oneAPI 2022.3) and the recommended high performance tensor
transpose (HPTT) library for maximum performance.  CTF can be regarded
as a DSL for tensor algebra that allows, for instance, to effortlessly
express a 2-mode tensor--vector contraction (i.e., $\mathcal{Y} =
\mathcal{A} \times_\pmf{2} \mathbf{x}$) within a single line of C++
code:
{\small
\begin{alltt}
  Y[\purple{"ijl"}] \gray{=} A[\purple{"ijkl"}]\gray{*}x[\purple{"k"}];
\end{alltt}
}
\noindent assuming that $\mathcal{A}$ (i.e., \texttt{A["ijkl"]}) is a
fourth-order tensor.  There are two important aspects about the
previous line of code that are worth discussing.  Firstly, both
tensors and the vector are implicitly distributed among different MPI
processes and, therefore, this operation is not exposed to the user
(CTF claims to minimize communication costs).  Secondly, each tensor
operation such as the one indicated above is susceptible\footnote{This
will depend on the contraction mode, the tensor splitting assumed by
CTF, and the number of MPI processes.} to assemble the resulting
partial subtensors $\mathcal{Y}^{(p)}$, in order to form the global
tensor $\mathcal{Y}$ (i.e., \texttt{Y["ijl"]}).  Therefore,
benchmarking such a CTF code example is similar to timing the
distributed contraction in addition to the tensor assembly.  This
global gathering incurs MPI communication, which can consume more
computational time than the tensor contraction itself.  This is
undesirable, especially in the case of the dHOPM$_3$, where the
partial subtensors resulting from each dTVC operation do remain
distributed throughout the entire execution (see
Algorithm~\ref{alg:HOPM}).

Figure~\ref{fig:HOPM-CTF} presents a strong scalability performance
comparison between our proposed dTVC algorithm and the CTF
implementation using up to 8 MPI processes (4 compute nodes) on MN4.
We show the normalized throughput measured in it/sp because,
contrarily to CTF, our implementation seeks one-dimensional optimal
splittings (see Fig.~\ref{fig:splitting})\pmf{. C}onsequently, order
7--8 tensors are contracted with 7 processes, while the ninth-order
tensor only exploits 5 processes.  Next, for each tensor in
Fig.~\ref{fig:HOPM-CTF}, the corresponding results are the aftermath
of averaging the normalized throughput values over all contraction
modes and all splitting dimensions ($d^2$ times) in the case of our
implementation.  Since CTF splits the tensors internally, the
associated results are only averaged over $d$ contraction modes.  In
both cases, the overhead related to the tensor assembly discussed
above is accounted for.

We shall start by focusing on the results of Fig.~\ref{fig:HOPM-CTF}
corresponding to a single MPI process \pmf{($p=1$)}, where only
shared-memory parallelism is available and there is no need to form a
global output tensor.  In this situation, our distributed
implementation is able to sustain more than 8~it/sp across all
tensors.  On the other hand, CTF only remains competitive for the
matrix case (order 2) and gradually loses performance, being more than
four times slower for the tenth-order tensor due to the subpar
performance of MKL's batch-strided kernels employed for the internal
contraction modes~\cite{Martinez2022}.  Getting back to the matrix
case using our proposed implementation (solid bars), it can be readily
seen that increasing the number of distributed processes has no
visible impact on performance: indeed, the \pmf{corresponding} results
are all above 80\% of the theoretical peak performance of the machine.
This is expected because the amount of memory being communicated using
one-dimensional splitting is reduced by a factor of $30623 \times$
with respect to the input tensor size (7.5~GB).  In contrast, much
smaller factors are associated with higher\pmf{-}order tensors (e.g.,
$8 \times$ for the tenth-order tensor) and thus larger communication
overheads as it can be inferred from the remaining solid bars of
Fig.~\ref{fig:HOPM-CTF}.  In this regard, CTF presents a rather
strange behaviour for third- and especially second-order input tensors
with throughput values well below 0.5 it/sp, thereby making it
unsuitable for distributed matrix--vector (or vector--matrix)
multiplication of large square matrices.  On the other hand, CTF
starts approaching and can even surpass the performance of our
implementation for higher\pmf{-}order tensors and more than four MPI
processes.  But such cases are largely dominated by the time spent
assembling the global output tensor, which can be more than one order
of magnitude higher than that of the TVC operation itself.  Indeed,
the bandwidth associated with $d>6$ and $p=8$ do not attain 10\% of
the peak performance and hence cannot be considered representative of
HPC scenarios.  Finally, it is worth mentioning that we have not
optimized neither the disjoint union, Eq.~\eqref{eq:TVC-k!=s}, nor the
global reduction, Eq.~\eqref{eq:TVC-k==s}, operations involved in the
formation of the global output tensor.

Figure~\ref{fig:HOPM-CTF} emphasises the criticality of the
distributed tensor--vector contraction and, particularly, its
associated global reconstruction process.  In this sense, the
convenience of expressing tensor contractions straightforwardly
through a DSL (via CTF, tiledArray, and similar libraries), can be
counterbalanced by the lack of performance arising from the continuous
assembly of the distributed output subtensors.  In the remainder of
this work, we will evaluate the performance of the dTVC algorithm
without reconstruction.  As previously discussed, this operation is
irrelevant in the dHOPM$_3$ because it naturally exploits the
distributive property of the tensor contraction.

\subsection{Distributed native tensor--vector contraction performance}\label{sec:dTVC-performance}
Table~\ref{tab:TVC} reports on the results of the dTVC ignoring the
global construction of the final tensor, which allow us to focus
exclusively on the computational kernel performance.  In particular,
we are concerned about the matriziced tensor's shape ---tall and
skinny vs.\ short and fat matrices--- on the contraction performance.
In addition to this, we propose the comparison of the looped approach
based on state-of-the-art MKL kernels against our \pmf{CPU-native}
implementation using two compilers\pmf{: icpx (Intel) and clang++
  (LLVM)}.  We remind the reader that results are averaged over all
contraction modes and all splitting dimensions and normalized by the
theoretical peak bandwidth corresponding to eight MPI processes on MN4
(order 7--9 tensors exploit lower MPI processes due to the
one-dimensional optimal splitting, as previously discussed).

It can be readily seen on Table~\ref{tab:TVC} that the looped dTVC
algorithm based on MKL's batch-strided kernel yields the worst
results, with relatively low average bandwidth and high standard
deviation rates.  The high variability of the results, except in the
matrix case, evidence the lack of mode-oblivious properties of this
implementation despite the fact that it is based upon a carefully
optimized kernel.  As demonstrated in Ref.~\cite{Pawlowski2019}, MKL
is more sensitive to the shape of the matricized tensor and, in the
case of concatenated small MVMs, LIBXSMM may be able to report
slightly better figures.  In any case, they are not expected to remain
competitive against those shown in the next two columns, which belong
to our \pmf{CPU-}native dTVC implementation using the same compilation
stack as well as the LLVM infrastructure.  Differences between
compilers are minimum and, in both cases, our proposed algorithm
achieves almost three quarters of the theoretical peak performance
(1024~GB/s) with a standard deviation below 10\%.  We remind that this
low value accounts for changes not only in the contraction mode, but
also in the splitting dimension which ultimately affects the shape of
the matricized form of the tensors.  Therefore, we can affirm that the
distributed strategy proposed in this work remains oblivious to these
two variables.

\begin{table}[tbh]
\centering
\caption{dTVC normalized, averaged bandwidth (measured as a percentage
  of the theoretical peak value) and corresponding unbiased, sample
  standard deviation percentage (within brackets) over all contraction
  modes and all splitting dimensions for the tensors of
  Table~\ref{tab:tensors}.  Results correspond to 8 MPI processes on
  MN4 and two TVC algorithms (looped and native).}
  {\small
  \begin{tabular}{cccc}
    \toprule
    Order & Looped (MKL) & Native (\pmf{icpx}) & Native (\pmf{clang++}) \\
    \midrule
    \phantom{1}2 & 76.4 (10.7) & 77.5 (15.6) & 80.7 (6.5) \\
    \phantom{1}3 & 65.7 (41.5) & 82.6  (1.9) & 83.1 (1.8) \\
    \phantom{1}4 & 58.3 (52.2) & 80.4  (5.7) & 81.1 (5.9) \\
    \phantom{1}5 & 52.8 (57.3) & 76.9  (6.3) & 77.6 (6.3) \\
    \phantom{1}6 & 47.0 (61.6) & 72.7  (8.3) & 73.3 (8.6) \\
    \phantom{1}7 & 39.7 (66.4) & 66.2  (9.2) & 66.6 (9.5) \\
    \phantom{1}8 & 38.8 (67.5) & 66.3  (9.0) & 66.8 (8.7) \\
    \phantom{1}9 & 40.7 (67.8) & 69.8  (9.2) & 70.1 (9.7) \\
              10 & 38.5 (70.2) & 70.6  (8.4) & 70.9 (9.3) \\
    \midrule
    Avg.         & 50.9 (55.0) & 73.7  (8.2) & 74.5 (7.4) \\
    \bottomrule
  \end{tabular}
  }
  \label{tab:TVC}
\end{table}
\begin{figure*}[tbh]
  \centering
  \begin{minipage}{0.32\textwidth}
  \subfloat[MN4]{
    \martinezFigLabel{0.9\textwidth}{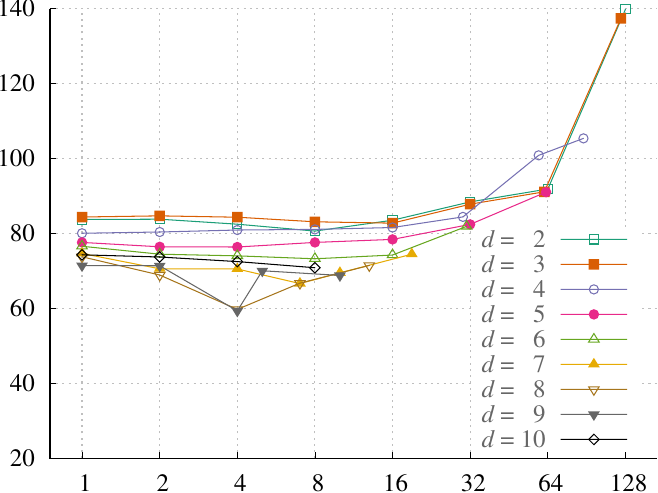}
    {MPI processes}{[--]}
    {Norm., avg. bandwidth}{[\%]}
  }
  \\
  \subfloat[MN4]{
    \martinezFigLabel{0.9\textwidth}{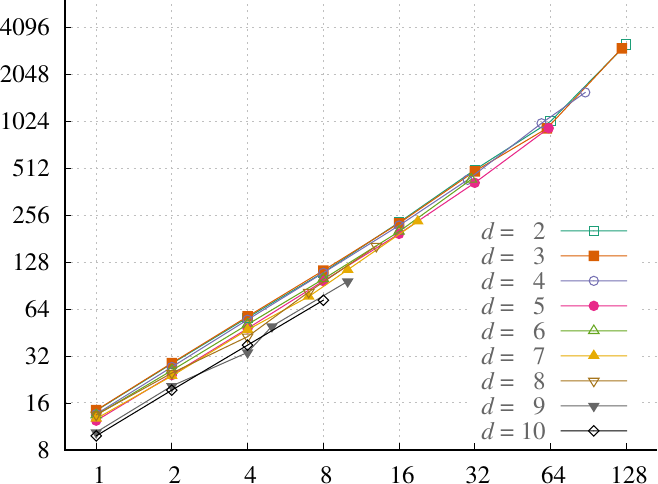}
    {MPI processes}{[--]}
    {Avg. throughput}{[it/s]}
  }
  \end{minipage}
\hspace{0.1cm}
  \begin{minipage}{0.32\textwidth}
  \subfloat[CTE-ARM]{
    \martinezFigLabel{0.9\textwidth}{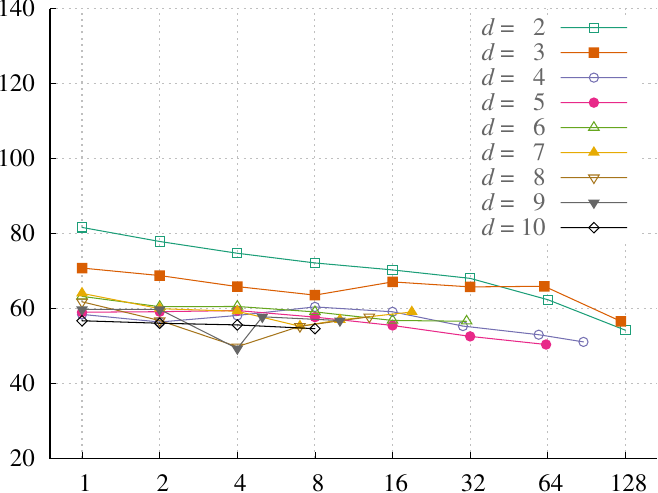}
    {MPI processes}{[--]}
    {Norm., avg. bandwidth}{[\%]}
  }
  \\
  \subfloat[CTE-ARM]{
    \martinezFigLabel{0.9\textwidth}{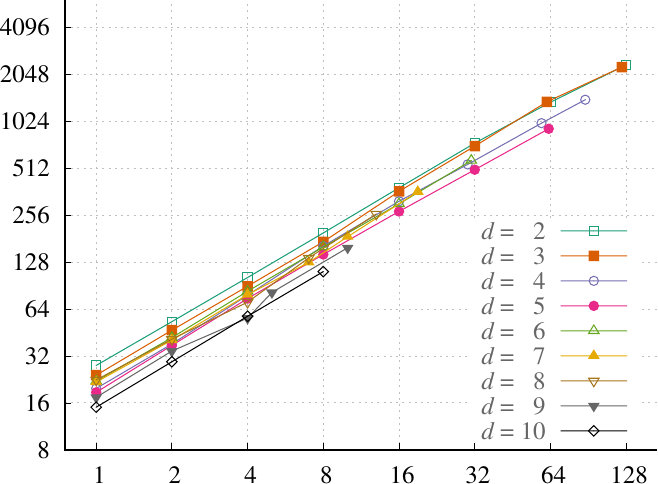}
    {MPI processes}{[--]}
    {Avg. throughput}{[it/s]}
  }
  \end{minipage}
\hspace{0.1cm}
  \begin{minipage}{0.32\textwidth}
  \subfloat[CTE-POWER]{
    \martinezFigLabel{0.9\textwidth}{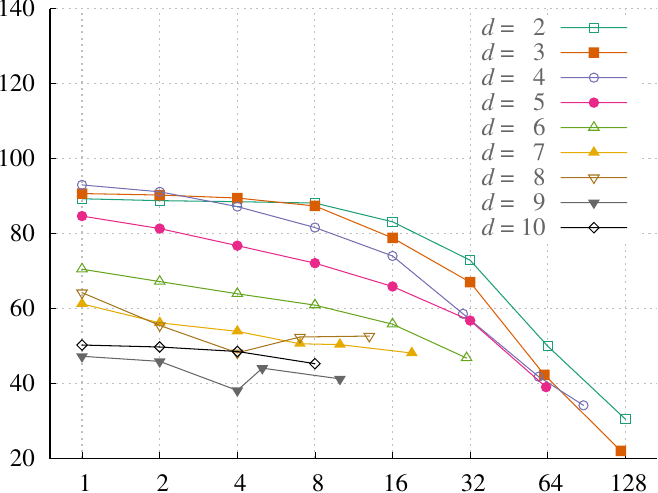}
    {MPI processes}{[--]}
    {Norm., avg. bandwidth}{[\%]}
  }
  \\
  \subfloat[CTE-POWER]{
    \martinezFigLabel{0.9\textwidth}{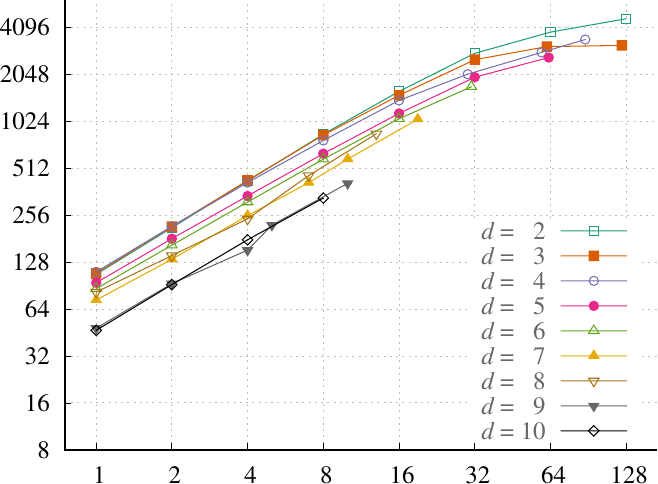}
    {MPI processes}{[--]}
    {Avg. throughput}{[it/s]}
  }
  \end{minipage}
  \hspace{0.1cm}
  \caption{dTVC normalized, averaged bandwidth and averaged throughput
    over all contraction modes and all splitting dimensions for the
    tensors of Table~\ref{tab:tensors}.  Strong scalability results
    using up to 128 MPI processes on (a)--(b) MN4, (c)--(d) CTE-ARM,
    and (e)--(f) CTE-POWER.}
  \label{fig:TVC}
\end{figure*}

Figure~\ref{fig:TVC}(a) shows that, in general, the averaged bandwidth
of all tensor contractions is between 70\% to 85\% of the theoretical
peak on MN4, which is in good agreement with the STREAM benchmark on a
single NUMA node of this architecture (81.6\%).  This rather small
performance variability of about 15\% is inevitable: for instance, not
all the tensors have the exact same memory footprint as reported in
Table~\ref{tab:tensors}.  With more than 16 MPI processes, the strong
scalability performance begins to increase slightly as the bandwidth
bottleneck moves from main memory to L3 cache.
Figure~\ref{fig:TVC}(b) shows absolute dTVC performance values in the
form of an averaged throughput measured in kernel invocations per
second (it/s).  We can see how the curves associated with each tensor
are very close to each other except for $d=9,10$ due to their larger
memory footprint (in relative terms: 6.7\% and 22.7\%, respectively).
\pmf{Note also that these two higher-order tensors cannot be computed
  with more than 10 and 8 MPI processes, respectively, when using
  one-dimensional splitting}.  We confirm the linear (and superlinear)
scalability already predicted by the bandwidth curves.  In the light
of the results gathered in Figs.~\ref{fig:TVC}(a)--(b), it is the
first time that a distributed tensor--vector contraction remains
oblivious to MPI splitting, contraction mode, and also tensor order,
provided that the tensor size does not vary significantly.

Figures~\ref{fig:TVC}(c)--(f) report on the dTVC performance for the
CTE-ARM and CTE-POWER systems\pmf{, respectively}.  We use the
\pmf{CPU-}native, TVC algorithm to benchmark the ARM architecture
while GPUs are evaluated against a looped TVC implementation making
use of cuBLAS.  Starting with the CPU case, Fig.~\ref{fig:TVC}(c)
shows slightly decreasing values of the normalized bandwidth,
especially for $d>3$, but they are overall within the range of 50\% to
70\% of the theoretical peak (on average, STREAM reports 60\% on a
single node).  The almost linear throughput curves of
Fig.~\ref{fig:TVC}(d) are a bit more spread than those of MN4 but,
once more, we can say that our novel implementation remains oblivious
to splitting, contraction mode and tensor order for this particular
architecture.  Finally, note that although CTE-ARM starts with better
absolute throughput values over MN4 thanks to its high\pmf{-}end
memory, its overall worse strong scalability yields subpar performance
with 128 MPI processes.

Figure~\ref{fig:TVC}(e) shows the dTVC performance on GPUs using
state-of-the-art cuBLAS kernels.  We observe a large variability going
from 90\% of the theoretical bandwidth peak to only 20\% (on average,
STREAM reports 61.8\% on a single graphics card).  Interestingly, only
the last two higher\pmf{-}order tensors report somewhat constant
bandwidth figures, but 10\% to 15\% below the reference STREAM
\pmf{benchmark} value.  This is expected since batched kernels do
incur some overhead while enqueueing consecutive multiplication
kernels within a unique function call.  This distributed GPU
implementation based on looped TVC lacks the obliviousness of its CPU
counterpart implementation.  In absolute terms, see
Fig.~\ref{fig:TVC}(f), the CTE-POWER attains maximum throughput values
around 4096~it/s, a quantity that is not much higher than the one
reported by a general-purpose architecture endowed with DDR4 memory.
It is worth highlighting that dTVC remains an embarrassingly parallel
application (the final disjoint union operation is discarded) with
zero MPI communication parts and, consequently, the scalability drop
in Fig.~\ref{fig:TVC}(f) is likely to be attributed to the rather low
device occupancy.  This supports our choice of one-dimensional tensor
splittings since very fine-grain parallelization tends to hamper
performance, especially on this type of accelerators.

\subsection{Distributed, native higher\pmf{-}order power method performance}
For the sake of conciseness, we only evaluate the performance of the
optimized dHOPM$_3$ (Algorithm~\ref{alg:HOPM}) and, therefore, naive
implementations are discarded in this work.  This choice is justified
because our results are ultimately compared against theoretical
bandwidth values in order to give an accurate representation on how
our distributed algorithm truly performs on a given architecture.  In
addition to this, tensors are split along their last dimension,
$\hat{s}=s/(d-1)=1$, to ensure minimal streamed memory and best
throughput (see Fig.~\ref{fig:touchMem}).

Table~\ref{tab:HOPM} \pmf{confirms the conclusions previously drawn in
  Section~\ref{sec:dHOPM-tasks} by demonstrating} how a genuine
data-flow execution based on either OpenMP tasks (OMP$_{\rm{tk}}$) or
OmpSs-2 tasks (OSS$_{\rm{tk}}$) can provide, on average, up to a 10\%
increase in memory bandwidth over a \pmf{canonical} strategy using
\pmf{fork-join,} parallel for constructs (OMP$_{\rm{fj}}$).
\pmf{OpenMP tasks are slightly faster than OmpSs-2 tasks in this
  particular case, which can be attributed to the design differences
  between these two runtimes.}  It is worth mentioning that these
moderate speedups are limited by the synchronous MPI collectives at
the end of each external iteration of dHOPM$_3$.  In cases where the
HOPM requires many iterations to reach converge ($i \gg 1$),
task-based relaxation techniques similar to those that have been put
in place in classical linear algebra iterative
algorithms~\cite{Martinez2023} could be employed to remove global
synchronizations and approximate the base of output vectors.  Such a
case is beyond the scope of this work, but represents a real scenario
where task-based parallelization could yield additional advantages.
\begin{table}[tbh]
\centering
\caption{dHOPM$_3$ normalized bandwidth (measured as a percentage of
  the theoretical peak value) over the last splitting dimension for
  the tensors of Table~\ref{tab:tensors}.  Results correspond to 8
  MPI processes on MN4 and three parallel implementations.}
  {\small
  \begin{tabular}{cccc}
    \toprule
    Order & OMP$_{\rm{fj}}$ & OMP$_{\rm{tk}}$ & OSS$_{\rm{tk}}$ \\
    \midrule
    \phantom{1}2 & 66.9 & 75.1 & 73.4 \\
    \phantom{1}3 & 63.8 & 83.7 & 80.0 \\
    \phantom{1}4 & 62.3 & 80.9 & 77.6 \\
    \phantom{1}5 & 59.7 & 74.1 & 71.8 \\
    \phantom{1}6 & 59.8 & 67.9 & 65.8 \\
    \phantom{1}7 & 54.2 & 59.8 & 58.4 \\
    \phantom{1}8 & 51.9 & 62.0 & 60.9 \\
    \phantom{1}9 & 60.2 & 63.3 & 62.9 \\
              10 & 59.6 & 70.9 & 69.5 \\
    \midrule
    Avg.         & 59.8 & 70.8 & 68.9 \\
    \bottomrule
  \end{tabular}
  }
  \label{tab:HOPM}
\end{table}
\begin{figure*}[tbh]
  \centering
  \begin{minipage}{0.32\textwidth}
  \subfloat[MN4]{
    \martinezFigLabel{0.9\textwidth}{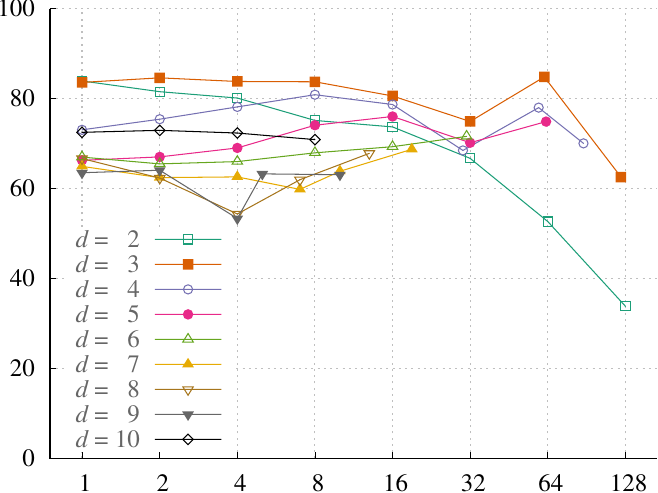}
    {MPI processes}{[--]}
    {Norm. bandwidth}{[\%]}
  }
  \\
  \subfloat[MN4]{
    \martinezFigLabel{0.9\textwidth}{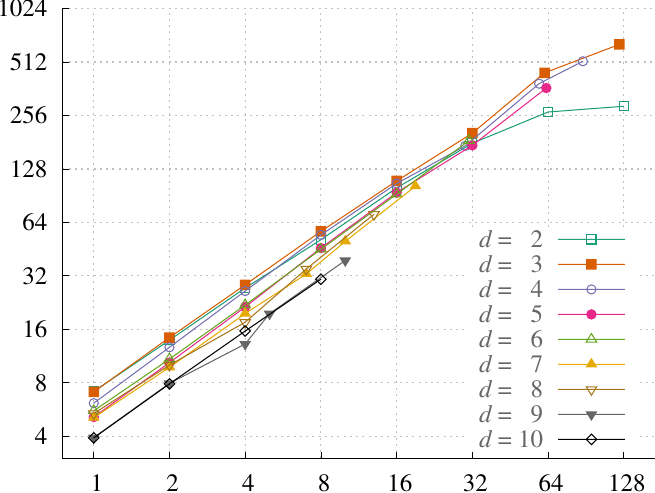}
    {MPI processes}{[--]}
    {Throughput}{[it/s]}
  }
  \end{minipage}
\hspace{0.1cm}
  \begin{minipage}{0.32\textwidth}
  \subfloat[CTE-ARM]{
    \martinezFigLabel{0.9\textwidth}{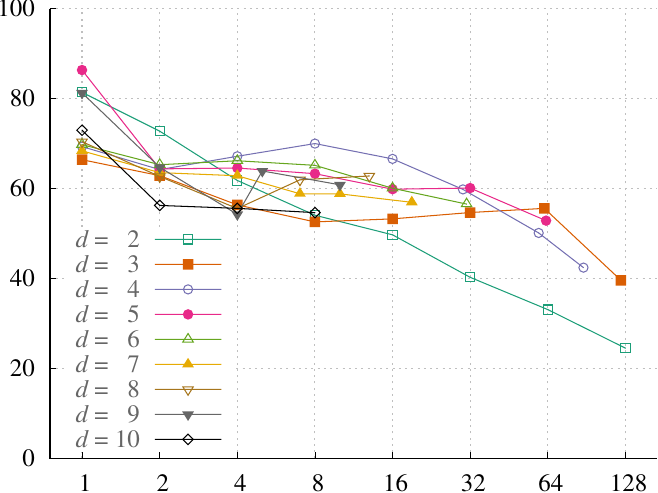}
    {MPI processes}{[--]}
    {Norm. bandwidth}{[\%]}
  }
  \\
  \subfloat[CTE-ARM]{
    \martinezFigLabel{0.9\textwidth}{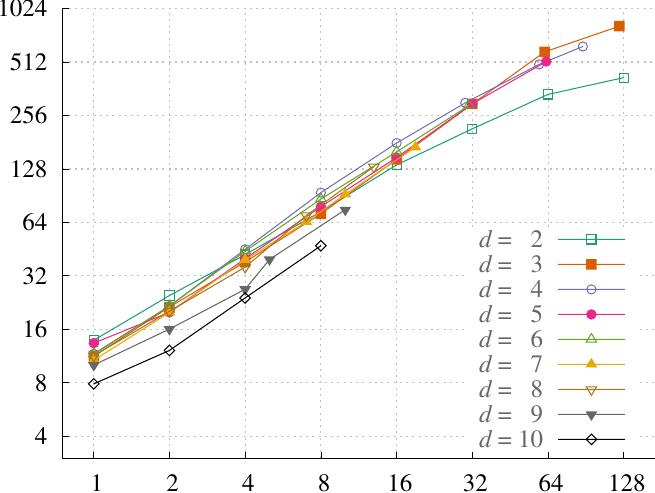}
    {MPI processes}{[--]}
    {Throughput}{[it/s]}
  }
  \end{minipage}
\hspace{0.1cm}
  \begin{minipage}{0.32\textwidth}
  \subfloat[CTE-POWER]{
    \martinezFigLabel{0.9\textwidth}{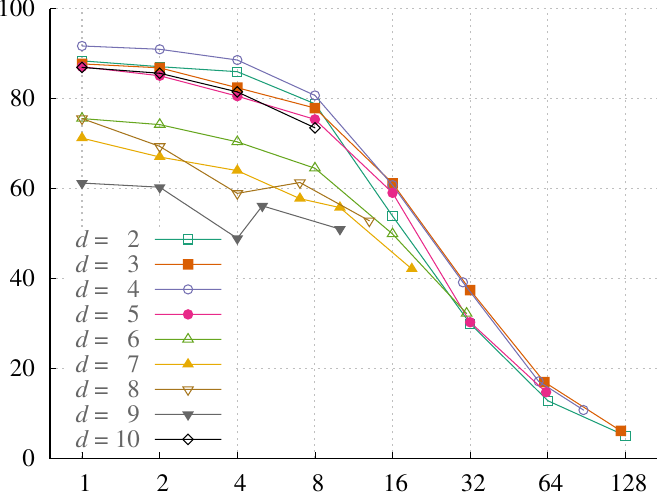}
    {MPI processes}{[--]}
    {Norm. bandwidth}{[\%]}
  }
  \\
  \subfloat[CTE-POWER]{
    \martinezFigLabel{0.9\textwidth}{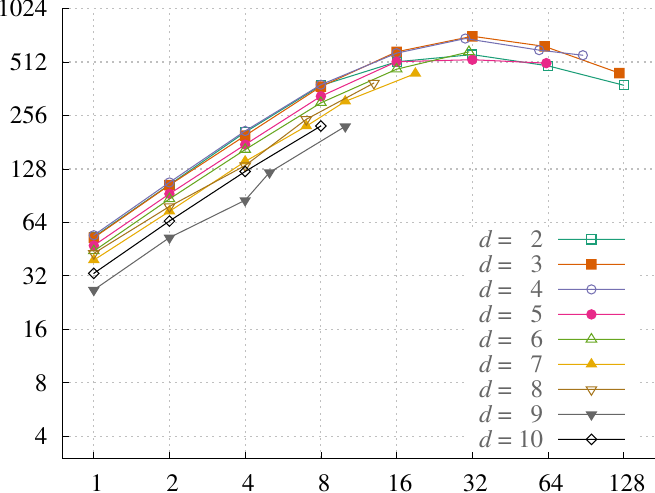}
    {MPI processes}{[--]}
    {Throughput}{[it/s]}
  }
  \end{minipage}
  \hspace{0.1cm}
  \caption{dHOPM$_3$ normalized bandwidth and throughput for the
    tensors of Table~\ref{tab:tensors} split along their last
    dimension.  Strong scalability results using up to 128 MPI
    processes on (a)--(b) MN4, (c)--(d) CTE-ARM, and (e)--(f)
    CTE-POWER.}
  \label{fig:HOPM}
\end{figure*}

Figure~\ref{fig:HOPM}(a) shows the normalized bandwidth of dHOPM$_3$
on MN4, which presents similarities with the dTVC bandwidth results
previously shown in Fig.~\ref{fig:TVC}(a).  The variability range is
expected to widen due to the continuous decrease of the tensor size
after each contraction but, in general, normalized values are between
60\% to 85\% most of the time.  Contrarily to dTVC, dHOPM$_3$ is not
an embarrassingly parallel application and, sometimes, the cost of
synchronous communications can be as important as computations as it
is the case with $d=2$ and $p>32$.  When looking at the absolute
throughput, see Fig.~\ref{fig:HOPM}(b), it starts at about 4--8~it/s
and reaches between 512--1024~it/s with 128 MPI processes for $d=3,4$
while it drops by a factor of two for the matrix case due to excessive
synchronization.  It can be readily seen that Figs.~\ref{fig:TVC}(b)
and \ref{fig:HOPM}(b) are reasonably equivalent with the particularity
that there exists a factor above two between them.  This is because
the computational time of dHOPM$_3$ is primarily dominated by two
contractions, \pmf{i.e.}\ $k=1$ for $j=0$ and $k=0$ for $j=1$, as
illustrated in Fig.~\ref{fig:HOPM-traces}(a).

Figures~\ref{fig:HOPM}(c)--(f) show the dHOPM$_3$ results on the other
two architectures equipped with high bandwidth memory.  In the case of
the CTE-ARM cluster, each contraction makes use of our
\pmf{CPU-}native TVC algorithm and no significant differences can be
appreciated between Figs.~\ref{fig:TVC}(c) and \ref{fig:HOPM}(c).
Nevertheless, it is true that, in the case of lower\pmf{-}order
tensors, dHOPM$_3$ bandwidth values are more sensible to the number of
processes due to synchronous communication.  A similar factor above $2
\times$ between dTVC and dHOPM$_3$ throughput\pmf{s} is found on
CTE-ARM.  Looking at the results obtained on these two CPU-based
architectures, it can be inferred that the dHOPM$_3$ algorithm
inherits the obliviousness of the native dTVC kernel it is built upon.

Figure~\ref{fig:HOPM}(e) shows the dHOPM$_3$ normalized bandwidth
under Volta GPUs.  Compared to the dTVC results of
Fig.~\ref{fig:TVC}(e), dHOPM$_3$ performance is better in the sense
that curves do remain closer to each other, especially with moderate
values of $p$.  This is in contrast to what is observed on the two
previous architectures.  As a reminder, dTVC results were averaged
over \emph{all} the splitting dimensions while dHOPM$_3$ figures are
only calculated for the \emph{last} splitting dimension for minimal
memory footprint.  For this reason, one can expect higher bandwidth
values for dHOPM$_3$ up to a certain number of processes; otherwise,
the relatively low GPU occupancy combined with the increasing
synchronization costs of collectives results in an important reduction
of bandwidth, with values as low as 10\% with 128 MPI processes.  It
is worth noting that these are indeed the best results one can achieve
in terms of communication: memory is transferred between devices
directly via the NCCL library that provides genuine nonblocking calls
to MPI functions from the CPU host.  Other communication approaches,
not shown here for the sake of brevity, always result\pmf{ed} in
subpar performance.  Finally, Fig.~\ref{fig:HOPM}(f) confirms that the
maximum throughput achieved on GPUs is above 512~it/s and does not
necessarily correspond to the largest number of processes.
Considering all the strong scalability results presented in this
section, none of the clusters equipped with high bandwidth memory is
able to surpass a general purpose supercomputer with regular DDR4
memory.  This is mainly attributed to the low\pmf{-}capacity HBM2
modules installed on these systems.

%
\begin{figure*}[tbh]
  \centering
  \subfloat[MN4]{
    \martinezFigLabel{0.9\textwidth}{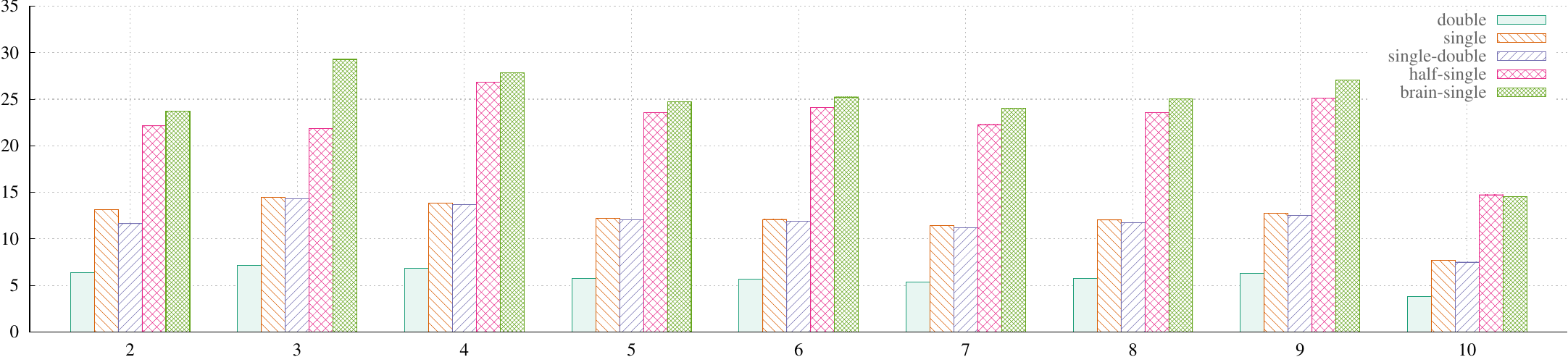}
    {Tensor order}{[--]}
    {Norm. throughput}{[it/sp]}
  }
\\
  \subfloat[CTE-ARM]{
    \martinezFigLabel{0.9\textwidth}{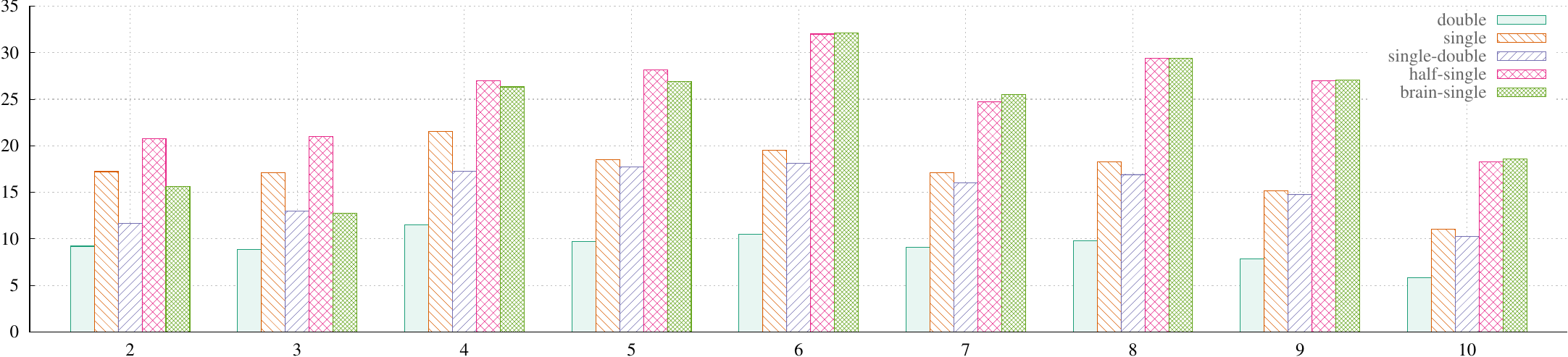}
    {Tensor order}{[--]}
    {Norm. throughput}{[it/sp]}
  }
  \caption{dHOPM$_3$ normalized throughput for the tensors of
    Table~\ref{tab:tensors} stored in different floating-point
    precision formats and split along their last dimension.  Results
    correspond to 8 MPI processes on (a) MN4 and (b) CTE-ARM.}
  \label{fig:HOPM-mix-prec}
\end{figure*}

\subsection{Mixed precision effects on performance}\label{sec:mixed-precision}
Since both dTVC and dHOPM remain memory bounded algorithms, mixed
precision can be employed to further increase their throughput.  In
fact, mixed precision is a very popular technique actively used in
deep learning applications~\cite{Gupta2015,Dhiraj2019}.  In this work,
low precision is used exclusively as a storage format and we promote
numbers to high precision, i.e.\ by doubling their bit representation,
just before computing them.  Therefore, every arithmetic operation,
besides accumulations, is done in high precision.  Calculated
quantities are then converted back to their original storage
precision.  It is worth noting that mixed precision also affects
communication and hence requires the development of ad-hoc MPI
functions that enable high\pmf{-}precision scalar and array sums while
keeping data transfers in low precision.

We consider three mixed precision formats: single-double, half-single,
and brain-single.  The half format stands for the IEEE 754-2008
binary16 standard while brain consists of a 16-bit truncation of the
binary32 type from the same standard.  While it is straightforward to
create a fast truncation algorithm for brain type conversions, we
employ a third-party library~\cite{Half22} for half type conversions
on MN4 exclusively because its CPUs do not provide native
\pmf{hardware} support.  Note also that conversions made in software
with this library inhibit the vectorization of some parts of our TVC
library, notably while storing the results from vector--matrix
multiplication kernels, vector updates (i.e., \texttt{axpby}
functions), and global array reductions via MPI.  As a result, the
entire application performance plummets making it unusable.  To
mitigate this, we introduce an intermediate array to cache
high\pmf{-}precision results.  The following code snippet
\pmf{extracted from Ref.~\cite{Martinez2024} and} corresponding to the
\texttt{axpby} function illustrates this:

%
{\small
\begin{alltt}
  \textbf{\teal{static const} \purple{intT}} unrollY \gray{=} simdObjSz\gray{*}simdObjSz;
         \textbf{\teal{const} \purple{intT}} nU      \gray{=} \textbf{\violet{MULTIPLE}}(n, unrollY);
  \textbf{\purple{objT}} wrk[unrollY] \textbf{\teal{__attribute__}} ((\textbf{\violet{aligned}}(cacheLn)));

  \textbf{\teal{const} \purple{lobjT}} \gray{*} \textbf{\teal{__restrict__}} p_x \gray{=} x;
        \textbf{\purple{lobjT}} \gray{*} \textbf{\teal{__restrict__}} p_y \gray{=} y;
  \textbf{\teal{for}} (\textbf{\purple{intT}} ui \gray{=} iZero; ui \gray{!=} nU; ui \gray{+=} unrollY) \{
    \textit{\brown{#pragma omp simd simdlen(simdObjSz)}}
    \textbf{\teal{for}} (\textbf{\purple{intT}} ri \gray{=} iZero; ri \gray{!=} unrollY; \gray{++}ri) \{
      wrk[ri] \gray{=} a\gray{*}\textbf{\violet{OBJT}}(p_x[ri]) \gray{+} b\gray{*}\textbf{\violet{OBJT}}(p_y[ri]);
    \}
    \textbf{\teal{for}} (\textbf{\purple{intT}} ri \gray{=} iZero; ri \gray{!=} unrollY; \gray{++}ri) \{
      p_y[ri] \gray{=} \textbf{\violet{LOBJT}}(wrk[ri]);
    \}
    p_x \gray{+=} unrollY; p_y \gray{+=} unrollY;
  \}
\end{alltt}
}

Caching is achieved by the array \texttt{wrk} that is aligned to a
cache line and can have an arbitrary size multiple of the SIMD vector
length known at compile time.  It can be readily seen that the main
loop is divided into one outer loop and two inner loops (the remainder
loop is omitted for simplicity).  The first inner loop can be
vectorized via OpenMP pragmas and accomplishes the actual computation,
$\alpha \bf{x} + \beta \bf{y}$, after promoting both input arrays to
higher precision (\texttt{OBJT}).  The second inner loop cannot be
vectorized due to the type conversion to low precision
(\texttt{LOBJT}) but the compiler is able to automatically unroll it
thereby achieving a throughput comparable to its single precision,
fully vectorized counterpart.

Figure~\ref{fig:HOPM-mix-prec} shows the absolute performance of the
dHOPM$_3$ algorithm for various floating-point precision formats on
two CPU architectures and eight processes.  Starting with MN4, the
throughput values corresponding to double precision are taken directly
from Fig.~\ref{fig:TVC}(b).  With single and single-double precision
the throughput is practically increased by a factor of two regardless
of the tensor order.  Brain-single precision doubles the speed of
single precision, yielding about $4 \times$ speedup with respect to
double precision.  The half-single precision results obtained with the
half library are quite competitive with the previous case demostrating
the advantages of caching.  Interestingly, there is a considerable
performance hit of $-34$\% for $d=3$, which makes for an effective
speedup of $3.1 \times$ over double precision.  In general, the matrix
case ($d=2$) yields slightly worse results for half-single and
brain-single precision, which is attributed to the global array
reduction involving tens of thousands of elements per process, all
carried out in mixed precision arithmetic.  By contrast, mixed
precision dTVC does not exhibit such behaviour since it does not rely
on collectives.  Generally, dTVC and dHOPM$_3$ mixed precision results
show similar trends, but the latter constitutes a more challenging
example due to its variety of kernels and synchronous collectives.

Figure~\ref{fig:HOPM-mix-prec}(b) shows inconclusive results on the
ARM architecture.  On the one hand, using single precision does not
necessarily double the performance as it was the case of MN4.  On the
other hand, and contrarily to what is observed in MN4, the
single-double results show hindered performance for lower\pmf{-}order
tensors ($d<4$).  Although the CPUs of CTE-ARM provide hardware
support for half precision floating points, half-single results
reflect a speedup below $2 \times$ w.r.t.\ single precision and about
$3 \times$ w.r.t.\ double precision.  Similarly to MN4, worse mixed
precision speedup figures are associated with lower\pmf{-}order
tensors due to the increasing cost of mixed precision collectives.
Finally, the brain-single results based upon a simple truncation made
in software show a performance similar to half-single mixed precision
arithmetic for most tensors ($d>3$).

Finally, we do not have mixed precision results on GPUs since, to the
best of our knowledge, NCCL does not provide mixed precision MPI
routines.  Moreover, CUDA kernels do not offer the degree of
flexiblity required to consistently integrate all the mixed precision
kernels needed by the distributed higher\pmf{-}order power method.  In
this regard, further work is needed to extend the support for mixed
precision in HPC software, including message-passing interface (MPI)
libraries.

\section{Conclusions and future work}\label{sec:conclusions}
This work has presented a distributed tensor--vector contraction
algorithm built on top of a \pmf{CPU-}native TVC shared-memory library
recently published.  It has also introduced a carefully optimized,
distributed higher\pmf{-}order power method, dHOPM$_3$, which takes
advantage of data-flow parallelization.  The performance of dTVC is
compromised when the splitting and contraction modes coincide and,
while this can be easily circumvented for one dTVC over a particular
contraction mode, it is unavoidable in the dHOPM algorithm where all
modes are constricted.  The analytical formulae derived from
hypersquare tensors demonstrate that the best parallel performance is
obtained when \pmf{the} splitting break\pmf{s} the contiguity of
tensor elements, in contrast to a naive approach that seeks to
preserve this property for simplicity.

Numerical experiments of dTVC and dHOPM$_3$ algorithms have been
carried out on three different architectures featuring CPUs, GPUs, and
high\pmf{-}end memory, using native and looped (cuBLAS-based)
implementations. On CPUs, we obtain dTVC bandwidth figures about 50\%
to 80\% of the theoretical peak using up to 128 MPI processes, which
are on par with those reported by the popular STREAM benchmark on a
single NUMA node.  In the case of dHOPM$_3$, figures are slightly
lower, but overall they follow the same trend and, in absolute terms,
the average throughput (kernel invocations per second) is, as
expected, roughly half of that of dTVC regardless of the tensor order.
On GPUs, the looped dTVC algorithm with state-of-the-art CUDA kernels
brings an excess of variability into the results and, more
importantly, evidences performance issues related to low device
occupancy and increasing synchronous communication.  In this regard,
general-purpose computers are still able to compete against newer
architectures equipped with low\pmf{-}capacity HBM or GPUs on strong
scalability scenarios.  Lastly, it has been proved that the use of
mixed precision kernels with our proposed caching technique is an
effective way of almost doubling or even quadrupling the throughput of
dTVC and dHOPM algorithms, although the real performance may remain
architecture-dependent.

Future work will focus on the development of a CUDA-based, native TVC
kernel for GPUs following the same philosophy adopted in our \pmf{CPU}
library.  This will enable beyond state-of-the-art dTVC performance,
and subsequently dHOPM performance, on these accelerators.  On the
other hand, it will be worth assessing the effects of multidimensional
tensor splitting and the incursion of tensor unfolding operations.
Analytical formulae for hypersquare tensors and numerical experiments
with fine-grain parallelization will shed light in the pros and cons
of increasing the number of splitting dimensions.  Finally, we will
seek to integrate this work within already existing frameworks,
e.g.\ by bringing dense multilinear algebra support to ALP.

\section*{Acknowledgements}\label{sec:acknowledgements}
This work was supported in part by MCIN/AEI/10.13039/ 501100011033 and
ESF/10.13039/501100004895 [grant number RYC2019-027592-I], and in part
by the HPC Technology Innovation Lab, a Barcelona Supercomputing
Center and Huawei research cooperation agreement (2020-2022).  This
work has also benefited financially from the Severo Ochoa Centre of
Excellence accreditation [grant number CEX2021-001148-S] funded by
MCIN/AEI. The Programming Models research group at BSC-UPC received
financial support from Departament de Recerca i Universitats de la
Generalitat de Catalunya [grant number 2021 SGR 01007].

\bibliographystyle{elsarticle-num-names}
\bibliography{bibliography.bib}





\end{document}